\documentclass[namedreferences]{kluwer}
\usepackage{epsfig} 
\usepackage{bm}

\def \bB {{\bf B}}

\def \bE {{\bf E}}

\def \bv {{\bf v}}

\def \del {{\bm \nabla}}

\def \div {{\bm \del} \cdot}
\def \curl {{\bm \del} \times}

\def \. {\cdot}
\def \x {\times}

\begin{document}
\begin{opening}
   \title{A LIFE of FUN PLAYING with SOLAR MAGNETIC FIELDS
     {\it (Special Historical Review)}}

   \author{E. R. \surname{PRIEST}}

   \institute{Mathematical Institute, University of St Andrews, 
              North Haugh, St Andrews, KY16 9SS, UK}

\begin{abstract}
This invited memoire describes my fortunate life, which has been enriched by meeting many wonderful people. The story starts at home and  university, and continues with accounts of St Andrews and trips to the USA, together with musings on the book {\it Solar MHD}.  The nature and results of collaborations with key people from abroad and with students is mentioned at length.  Finally, other important aspects of my life are mentioned briefly before wrapping up.  
\end{abstract}

\end{opening}
\section{Introduction}\label{sec1}

As I look back over my career, the main feeling is one of gratitude for having been blessed by a wonderfully supportive family, having met so many marvellous colleagues and scientific friends, and having been inspired by such interesting research. 

When thinking of life as a university teacher and researcher, I recall the saying of Confucius ``Choose the profession you love, and you won't have to work a day in your life".  For me, it has been an ideal job, with little sense of drudgery or stress, and with every week bringing new ideas and interesting experiences.  Teaching and research have always been equally important, and I have put just as much energy, enthusiasm and ingenuity into communicating to students as into wrestling with new research ideas.  Indeed, when hunting for a job, it was a positive decision to go into university teaching rather than apply to a research institute because I wanted to give back to others something of what I had learned. Also, on application forms, I have invariably described my profession as ``university teacher".

Three years ago, I ``retired", which has meant no salary and no responsibility in terms of applying for grants or being in charge of the research group, but otherwise there has been little change -- I still have an office and the same work hours, and continue to interact and collaborate with students and staff in the group and elsewhere.  In other words, I am having a great time enjoying myself. Indeed, since recently finishing a rewrite my old book {\it Solar MHD}, I have lots of ideas longing to be worked out.

For me, the following aspects are important:  

{\bf *} being open to new ideas and recognising that my ideas will change, which in turn reveals how little I personally know and understand (for example, when briefly sharing a room in Boulder in 1971 with Jan Stenflo, he told me he had inferred the presence of kilogauss fields in the quiet Sun -- but I thought he was crazy!);  

{\bf *} respecting other people's ideas or approaches when they differ from my own; 

{\bf *} encouraging a sense of community, where all are valued and respected;

{\bf *} keeping a balance between my own areas of speciality and a general overview of the development of solar physics as a whole or indeed of science as a whole;  

{\bf *} encouraging, valuing and helping young researchers develop their ideas and careers;

{\bf *} and maintaining a critical but high-quality attitude to research; when listening to a talk, I have always encouraged young researchers to be critical of the ideas but to be generous to the person, and to ask questions with a view to helping understanding rather than showing how clever you are or how stupid the speaker is.

\section{Early Days}\label{sec2}

I had a simple, humble home background, and was the first member of my family to attend university.  I grew up in the west of Birmingham, UK, with a hard-working father who instilled many basic values, such as that people should be able to think, believe and express what they liked in a civilised society, provided they agreed to live in peace and show respect for other members. He used a wide range of colourful local expressions such as ``stone the crows" and ``God bless the prince of Wales".  My mother was and is (at the age of 97) a caring parent with little education due to childhood ill-health, but a bright mind -- when I visit her today she does the crossword each morning and invariably asks for a game of scrabble in the evening.  My sister has my mother's sense of love and care, and my brother has been a close friend who has given us invaluable practical help throughout our life in St Andrews.

I attended King Edward VI School, Five Ways, in Birmingham, with a set of dedicated and inspirational teachers (such as Ken Thomas), who communicated a deep love of their subjects, from the richness of the English language to the joy of French and the fascination of mathematics or physics. This opened my eyes to how many intriguing topics there are to learn about in our short lives. Indeed, one of my habits whenever I am in St Andrews is always to attend an inaugural lecture from a new professor, no matter what the topic, and I have invariably been fascinated and learnt something new and interesting.

\begin{figure}
   \centering
   \includegraphics[width = 10cm]{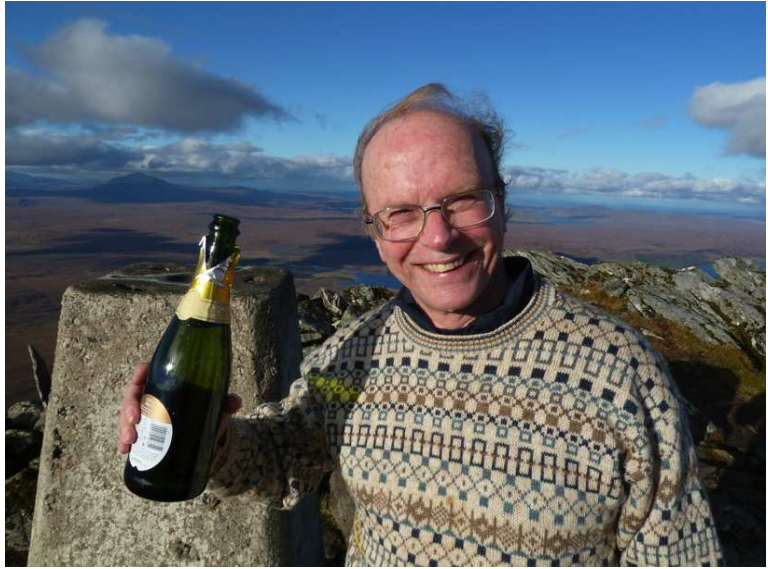}
   \caption{Celebrating at the summit of my last Munro with my son, Matthew.}\label{fig1}
\end{figure}
At school I also learnt the joy of sport (swimming, rugby and athletics), which has stayed with me throughout my life.  I still cycle to and from work every day and go to a circuit or aerobics session in the gym twice a week.  In Scotland, for twenty years I have been climbing ``Munros", i.e., mountains over 3000 feet in height, which has taken me to some incredibly beautiful and memorable locations, many of them quite remote. Each of them gives you a really good workout, and some of them can take up to 12 hours to climb.  It was a thrill last year to climb the last of the 282  Munros with my son Matthew, who brought out a bottle of champagne at the summit (Figure \ref{fig1}).  Three people soon joined us at the top and we offered them a glass of bubbly, but they declined, saying they were teetotal!   So Matt and I had to finish the bottle ourselves, and, needless to say, the walk down the hill was rather less straight than the walk up.

Another hobby that began at school was a love of music.  Indeed, my mum saved up ten shillings (50 pence)  per month to pay for a piano for me.  I have always sang in a large choir, usually as one of the quiet unconfident people at the back, but after turning 60 I decided to have singing lessons.  What a revelation! It was like going up into the loft, picking up a rusty old instrument and finding what amazing sounds could be made with it.  Since then, I sing for half an hour each morning before work, and my voice has developed a lot, both in terms of tone, range, control and sight-reading.  I sing in two choirs each week, one at church and the other a town-gown choir called St Andrews Chorus.  This is led by an inspirational conductor who is head of music at the university, and I have been the president of the Chorus for the past 7 years, during which time the size has almost doubled to over 160 members, half being students and half townsfolk.

\section{At University -- Nottingham, Leeds and Cowling}\label{sec3}

In my last two years at school,  two-thirds of the time was spent in the amazing world of mathematics, and the remainder was studying physics.  This gave me an excellent grounding in these subjects, but was more specialised than the Scottish system, which I now strongly prefer.   One of my father's sayings was that ``your word is your bond", and that in the Midlands if you promised something then you must carry it out.  So, when Cambridge University offered me an undergraduate place a week after I had accepted one at Nottingham, I clearly had to turn down Cambridge, saying that I was already going elsewhere.  In a parallel universe where I had made a less moral decision, I doubt that I could have been any happier with my journey through space and time.   

Life at Nottingham (1962--65) was full of positive experiences,  learning mathematics and physics as well as developing diverse interests in sports,  music and Christianity. I do not remember studying any evening during my 3 years, since I could understand most of the lectures and so just brushed them up and read around the topics in the afternoons. While there, I became fascinated by the fact that fluid mechanics has the vorticity being the curl of velocity, while electromagnetism has current being the curl of the magnetic field. I was unsure whether to do a PhD on the one or the other, and then I discovered that a relatively new subject, magnetohydrodynamics (MHD for short), included both vorticity and current, and so that was the subject for me.  

I also discovered that the UK's expert on MHD and one of the founders of the subject was T.G.Cowling at Leeds, so I wrote to him and asked if he would supervise me for a PhD.  He agreed, but the standard procedure at the time was first to do the MSc course in Applied Mathematics and so I started research  a year later (1966).  The MSc (1965--66) was full of matched asymptotic techniques and examples such as boundary layers, which were all the rage at the time, and it had marvellous lectures from Cowling himself and others such as John Brindley.

\begin{figure}
   \centering
   \includegraphics[width = 6cm]{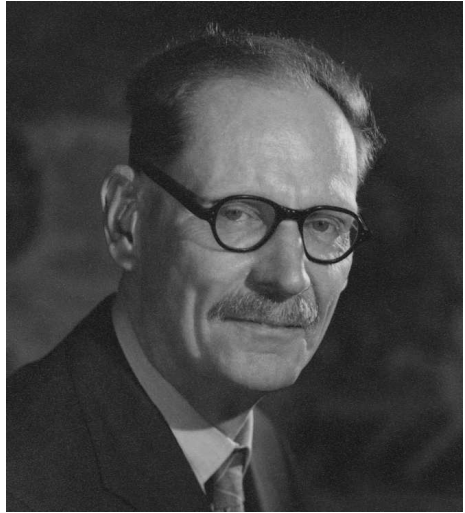}
   \caption{T.G.Cowling (a kindly man of fearsome intellect and a real privilege to have known) courtesy of the Godfrey Argent Studio.}\label{fig2}
\end{figure}
Cowling (1906--1990) was an eminent figure with high critical standards, who was a Fellow of the Royal Society, president of the Royal Astronomical Society and winner of its Gold Medal, but he was a kind and humble man with a deep sense of justice and integrity (Figure \ref{fig2}). As the British astronomer Leon Mestel has commented,  ``With his death the Royal Society has lost one of its most distinguished and well-loved fellows... I shall always think of him as representing the best in the British Puritan tradition". Cowling wrote about 70 research papers on topics ranging from the theory of stellar structure and nonradial oscillations to dynamo theory, the structure of sunspots and electrical conductivity.  In his autobiography he comments over-generously: ``Just before I retired in 1970, I had as a PhD student Eric Priest. It was only after some hesitation that I took him on, since I felt that I was getting too far out of touch with recent developments. His cheerful enthusiasm more than compensated for my deficiencies." (I certainly did not notice any deficiencies!)  Each week he would spend a couple of hours with me.  I would knock on his secretary's door and be shown into his room, where he would greet me with ``Good morning, Priest, how are you today?" ``I am fine, sir, how are you" like a Japanese tea ceremony.  I would sit down at his desk and show him my workings for the week, and it would invariably end up with ``Yes, very good, but ...".  We would discuss where to go next with the work, and then he would lean back and say ``Is there anything else you would like to ask me?", at which point I could ask about a physical or mathematical topic or about the history or development of ideas or about individual people or how he had made a discovery.

I remember asking him on one occasion whether mathematics or physical intuition was more important, to which he pondered for a while before replying ``Well, both are important, since you need to ensure both a mathematical  and physical understanding in any problem.  But, if they disagree, I would recommend siding with physics." He also described what it was like doing research in the early days, how, for example, he would write a letter to Biermann about some theoretical topic and would spend several days carefully crafting the letter, only to wait perhaps a month or two for a reply. In response to a query of mine about what Alfv\'en was like, he replied sadly that he had once written to him about a topic on which they disagreed, ending with a hope that, in spite of disagreeing they could remain good friends, to which Alfv\'en responded:  ``Nobody who disagrees with me can possibly remain a friend". However, in spite of criticising Alfv\'en's work on several occasions, he also generously commented ``Alfv\'en's best ideas are very good indeed". On another occasion, I asked how he came to discover his anti-dynamo theorem in the early 1930's, to which he described how he had been trying to calculate the internal magnetic field of a star by solving the equations {\it numerically}, by hand: -- i.e., using finite differences on a grid of, say, 20 by 20 in an axisymmetric plane. He tried lengthy calculations several times, but on each occasion the method failed near an O-line. Eventually, he realised that perhaps the equations possessed no solution, and so he played with the basic equations until he understood why.

Working with Cowling (as his last research student) was a real pleasure. I learnt so much from him. He gave me a desire to work hard and try to produce work of the highest quality, to master the relevant mathematical techniques and to aim to develop deep physical intuition.  He also stressed the importance of asking questions and trying to probe to the core of a problem, never being satisfied with a superficial answer.

\section{St Andrews}\label{sec4}

\subsection{Early Years and Lecturing}

After doing two years towards my PhD, my funding ran out, since students were supported for only 3 years, including the one-year MSc course.  So I started to apply for jobs.  I was fortunate to be given a permanent (tenured) lectureship at St Andrews at the age of 24 in the Applied Mathematics Department from September 1968, in spite of having no publications -- how things have changed!

I arrived at St Andrews by train (along a line that no longer exists) from what was then Leuchars {\it junction}. As the train wound its way along the Eden estuary past the sand dunes and golf courses, and the spires of the ancient town appeared ahead, it seemed like entering a magical world -- like Narnia or Hogwarts.  This was the oldest university in Scotland, founded between 1411 and 1413 (so we have just celebrated our 600th birthday) and it has many historical buildings as well as one of the highest standards for academic research in the UK.  

St Andrews did seem rather remote and so I expected to move on to somewhere different after a couple of years.  However, like many before me, I fell in love with the place (and also with my wife Clare) and so decided to stay.  It has been fun gradually  building up a research group with outstanding members, and one key aspect that has kept my research alive has been going away each summer (usually to the USA) for a {\it scientific retreat}, where I could recharge my batteries and develop new ideas.

My first year as a lecturer was busy -- apart from doing the third year of my PhD and writing up, I had a full lecturing load and was learning how to teach.  At first, I was petrified to stand in front of a hundred students, but now I love it and relish the challenge of putting things over in a clear manner. One of my habits was to give the students a ``joke-break" in the middle of a one-hour lecture, during which I would practice my range of accents; this was not only for fun, since I realised the students could not concentrate for a full hour and so this gave a natural break, after which both the students and I could restart with new enthusiasm.  Crafting a good lecture was non-trivial, since I always wanted the students to leave at the end having learnt something new, rather than just having transferred the lecture notes from my book to theirs without going through either their minds or mine.  I tried to insert a number of ``disclosure points" in each lecture, where we would both stop writing and I would help the students go through a crucial part of an argument or realise something really significant. 

In the early days, the job as a lecturer seemed such fun (and has remained so), without the slightest pressure (by comparison with what I observe for young researchers these days).  We seemed to have endless time to think, with little administration, and no pressure to publish more than one or two papers per year.  I used to play chess with Bernie Roberts some afternoons and we used to go to coffee with colleagues from many other arts and science departments. Frequently, we would go to hear a seminar in philosophy or some other department, as the mood took us.

\subsection{Applied Mathematics}

In St Andrews, I teamed up with Jeff Sanderson and did some research on collisionless shock waves, solving the coupled Maxwell-Vlasov system for the effect of gradients on micro-instabilities such as Bernstein and ion-acoustic instabilities, and so my first few papers were on this topic. It was a couple of years later that Cowling suggested I write up one of the topics from my thesis.  

I am basically an applied mathematician with a knowledge of various mathematical techniques, which can be applied to whatever field I like, but it is mainly the solar magnetic field that has captivated me, with so many rich ideas and problems longing to be tackled and inspired by a series of new observations. I have always said that, if I become bored with solar physics or run out of ideas, then I will apply my wares elsewhere, but there is no sign of that happening anytime soon.  Occasionally, I have turned to astrophysical problems, especially with Jean Heyvaerts, with forays into  astrophysical jets, accretion discs or star formation, and I am very much aware that new solar knowledge is asking to be applied elsewhere in the universe.  

The field of applied mathematics is mainly a UK phenomenon, so that, if I had been born elsewhere, I may very well have been in a department of physics or astronomy.  It grew greatly in the 1950s and 60s as new departments focusing on fluid mechanics were born with an aim to understand fluid flows and turbulence.  Since then topics of numerical analysis, MHD, plasma physics and biological fluids have been added. Much of applied mathematics is really theoretical physics in the broad sense, since the mathematical techniques are being applied to physical problems.  So theoretical plasma physics and solar physics are often studied in Applied Mathematics departments in the UK, whereas other parts of physics (such as quantum physics and relativity) are invariably located in Theoretical Physics departments, but I have often felt that the distinction between the two types of department is artificial, since we have essentially the same philosophy and use similar mathematical tools.

\subsection{The Solar MHD Group}

Three years after arriving in St Andrews, I had a year's study leave in the USA and during that year, Bernie Roberts was appointed as my temporary replacement. Fortunately, on my return in 1972, a permanent lectureship was available and it was great that Bernie gained that position, so that the solar MHD group was born.  Bernie has always been a highly imaginative researcher with many original ideas, which I greatly admire, and he soon became a world expert on MHD waves.  We have always been good friends and have followed closely each other's research topics and supported each other. It became natural for him to focus on one growth area in MHD (namely waves), while I focused on others (reconnection, equilibria and instabilities), but we have occasionally dabbled in each other's, as when he discussed Petschek's mechanism (Roberts and Priest, 1975) or I discussed phase mixing of waves (Heyvaerts and Priest, 1983).   

\begin{figure}
   \centering
   \includegraphics[width = 10cm]{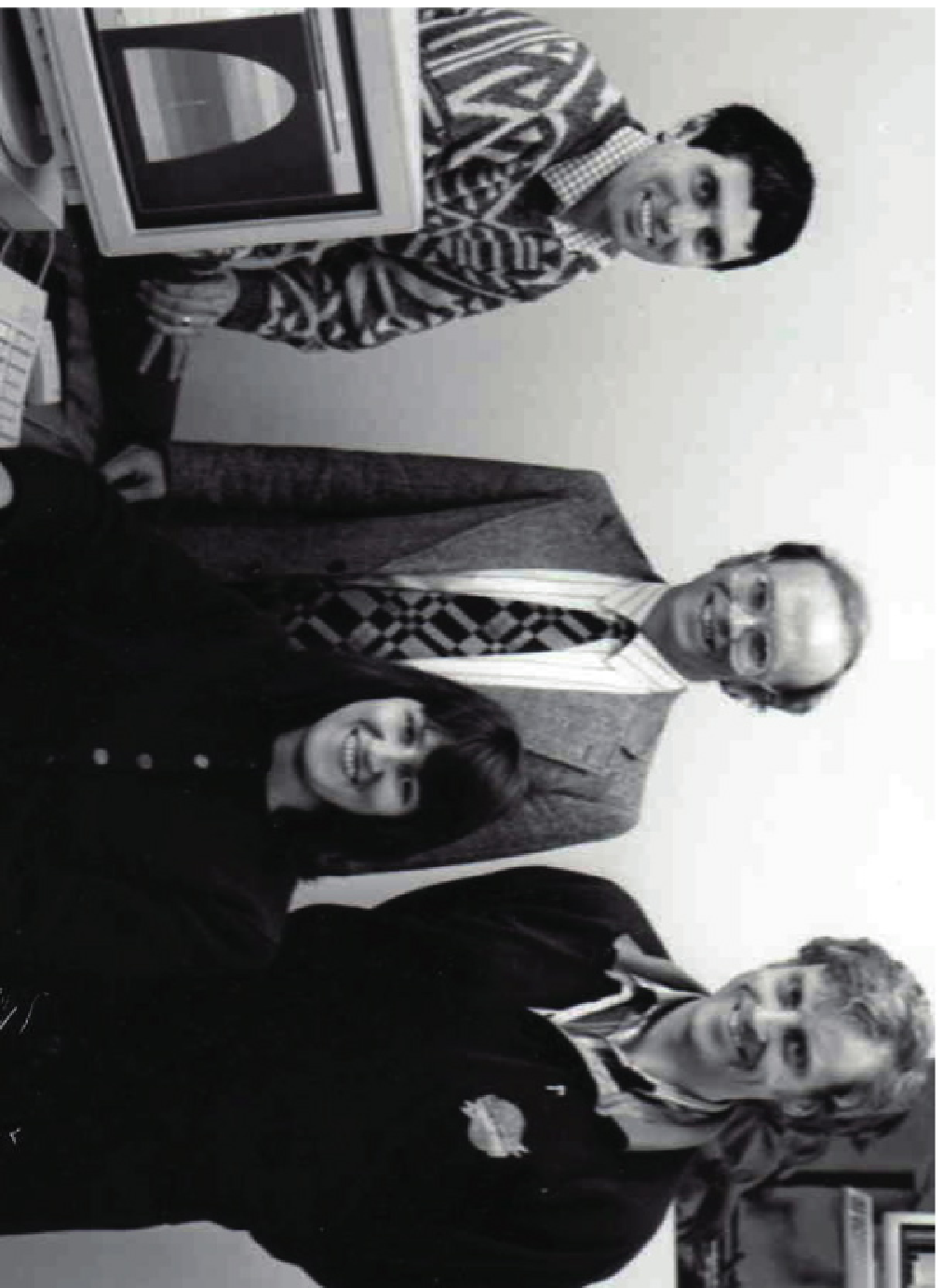}
   \includegraphics[width = 10cm]{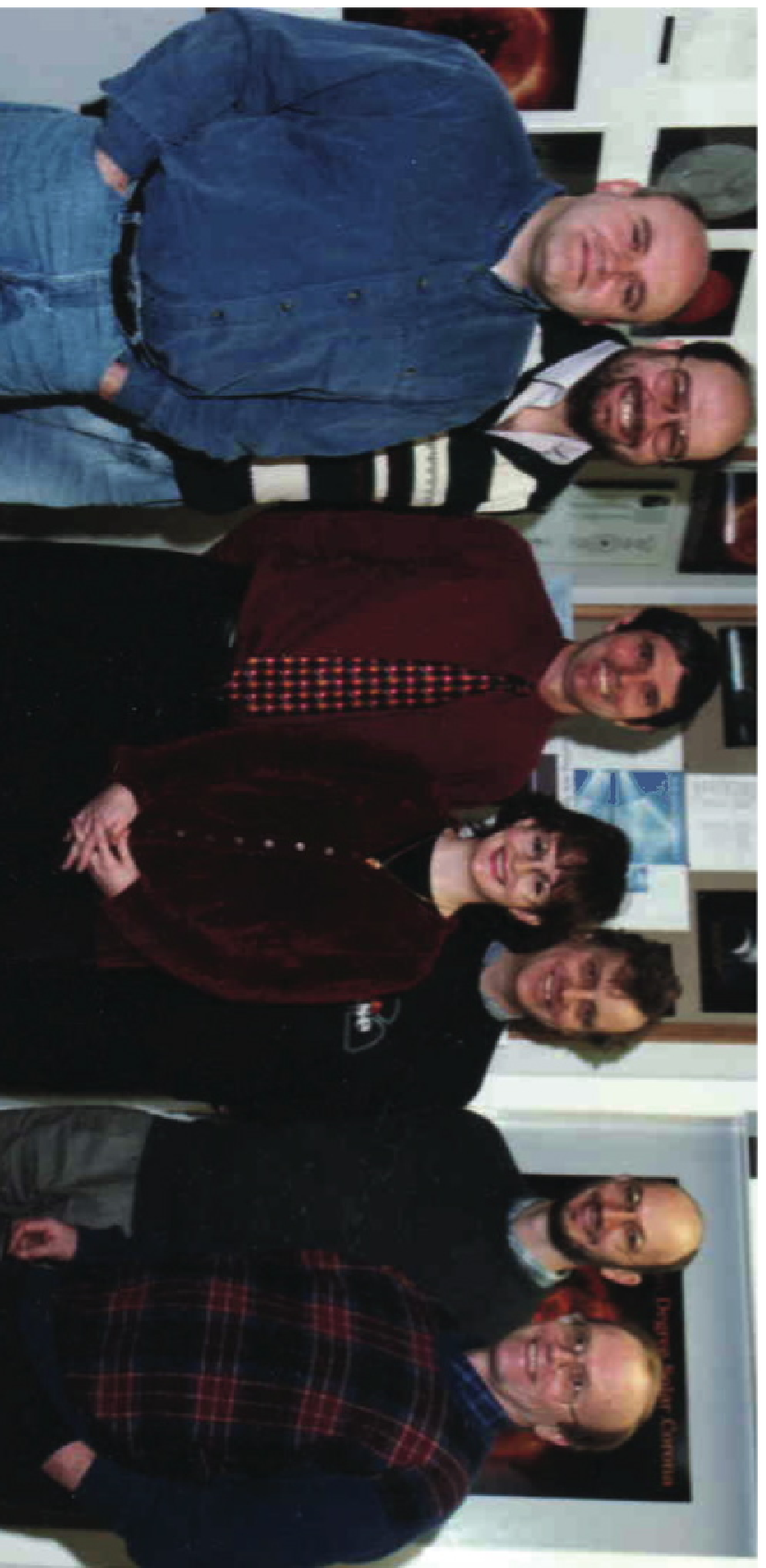}
   \includegraphics[width = 10cm]{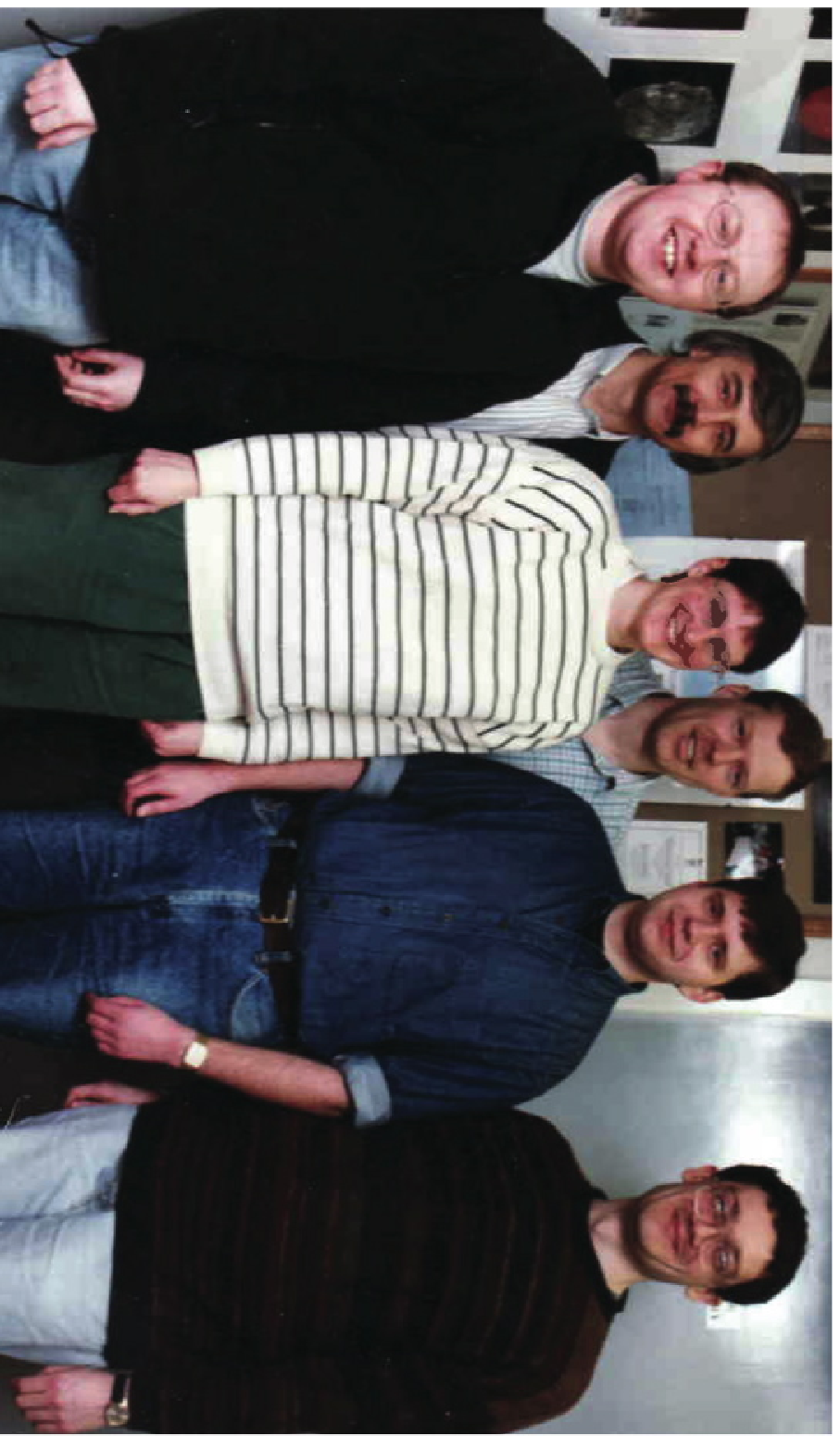}
   \caption{The Solar MHD Group in St Andrews: (a) the staff in April 1990 -- Bernie, myself, Hania (senior programmer) and Alan; (b) the staff in March 1999 -- Tony Arber (now prof. in Warwick), Thomas, Bernie, Gill (our fabulous secretary), Alan, Andrew and myself; (c) postdocs in March 1999 -- Robert Walsh (now reader in Preston), Michael Ruderman (now prof. in Sheffield), Clare (now prof.), Klaus Galsgaard (now associate prof. in Copenhagen), Duncan (likely soon to be prof.), and Aaron Longbottom (computational genius).}\label{fig3}
\end{figure}

I have always found it sad when an active researcher retires and leaves behind no permanent staff to continue the field at a university, and so I have worked hard to build up an enduring legacy at St Andrews, but this would not have worked without researchers of high international quality.  The solar MHD group (Figure \ref{fig3}) developed slowly from its early beginnings with the first two research students in 1973, the first postdoc in 1979--81 (Alan Hood), and the first extra permanent lecturer in 1984 (Alan Hood). This meant we could develop three subgroups working on MHD waves, reconnection and instabilities, each led by a lecturer. Later, Andrew Wright joined us in 1992, adding an expertise in magnetospheric physics, and Thomas Neukirch came in 1995 as a postdoc from the fabulous stable of Karl Schindler (one of my heroes) with wide interests in equilibria and kinetic theory. In the UK, there is a scheme of highly competitive research council 5-year awards called Advanced Fellowships (AFs) (and a similar scheme of Royal Society University Research Fellowships (URFs)).  Only a few of them are awarded in the whole of astronomy each year, and they represent recognition of international excellence and future expected research leadership. Within a year of Thomas being awarded an AF for the period 1997--02, the University naturally agreed to guarantee him a permanent position when the five years of funding came to an end. 

A new growth of the group occurred about ten years ago when Clare Parnell (2003) was appointed an AF (2002--2008), followed by Duncan Mackay (2004--2009), while Ineke De Moortel was appointed a URF (2004--13) and so was Vasilis Archontis (2011--2016).  They were each also guaranteed permanent lectureships in the group within a year of being appointed to their fellowships. 

It is clear that, with Clare, Duncan, Ineke and Vasilis in place, the future of the group is in excellent hands when the rest of us fade into retirement.   They are each high-quality original thinkers, with outstanding reputations in their own areas, and have research students and/or postdocs helping their work.  Front-rank research is being done by Clare on coronal heating and reconnection, by Duncan on prominences and flux transport, by Ineke on waves and by Vasilis on flux emergence. Indeed, each permanent member of the group is highly independent in their thinking and research, but communicates effectively with other members of the group and so benefits greatly from the lively and supportive research environment, where ideas are freely offered to help each member flourish.

Another joy has been to see a fantastic new solar MHD group develop in nearby Dundee University, based around Gunnar Hornig.   Gunnar is a brilliant scientist and a good friend, who understands 3D reconnection at a much deeper level than me and from whom I have learnt much of what I know about the subject. He completed his PhD with Karl Schindler in Bochum in 1997 and then led a temporary group there on topological fluid dynamics from 1998 to 2004. He visited us for 3 months in 2002 and then was an 18-month visiting researcher in St Andrews from April 2004 and given a lectureship in Dundee from 2005.  This has been supplemented by permanent positions for two fantastic former research students of mine, David Pontin from 2007 and Antonia Wilmot-Smith from 2010, who are revolutionising our understanding of magnetic topology and reconnection.

\subsection{Resarch Students and Postdocs}

Bernie and I have had an average of one new research student per year between us, and have been fortunate in their high quality. Indeed, I have usually found that after one or two years, they know more than me about the topic and so it evolves into a genuine collaboration.  We have always regarded a PhD as a training in research, the aim being to teach the student all the skills required for doing independent research, including selecting a topic, working it out, writing it up and giving effective talks about it.  As such, having a research student is a huge responsibility, both in selecting a really good topic, valuing them, having weekly in-depth discussion and helping them develop their talents and career.  For me, the absolute maximum to be supervising effectively at any one time is three.  So how do you define a PhD?  I adopt a pragmatic definition of what a reasonable student, working reasonably hard (i.e., putting in the hours and treating it as a 9.0am to 5.0pm activity like a job) can achieve over three years (which is the normal length for PhD funding in UK), and indeed the vast majority of students in St Andrews have completed their PhD in 3 years plus perhaps one or two months.  

The British PhD has the advantage of being completed only 3 years after the first degree and having a very high completion rate (I can remember only one student not finishing their PhD in the group in the past 40 years), so that a student is still young enough to embark on a completely different career.  However, it is highly specialised, and so most postdocs work hard at continuing to broaden their knowledge. Clearly, there are pros and cons when comparing the American or continental systems with the British.

My first few research students were Ed Smith (1973--76) on thermal conduction in prominences [he went on to make models of conduction in barley kernels at the agricultural research institute of Scotland],  Tim Tur (1973--76) on current sheet formation [he became head of maths at a sixth form college], Alec Milne (1975--78) on prominences [he runs his own scientific consultancy firm].  Amongst those that followed, particularly notable were Alan Hood (1976--79), Peter Cargill (1978--81), Philippa Browning (1980--83), Moira Jardine (1985--88), Clare Parnell (1991--94),  Duncan Mackay (1994--97), Dan Brown (1996--99), David Pontin (2001--2004) and Antonia Wilmot-Smith (2004--07).  The first six are now full professors in UK universities and the last four I expect to obtain such promotion in future.  I am sure I have learnt much more from this brilliant group of researchers than they may have learnt from me -- how fortunate I have been to be associated with them.

It gives a lot of pleasure to see how many former postdocs from St Andrews are now flourishing elsewhere as leading members of our field, such as Terry Forbes (1980--84), Peter Cargill (1981--82), Philippa Browning (1983--85), Marco Velli (1985--89), Mitch Berger (1986--89), Sami Solanki (1987--89), Tahar Amari (1988--90), Valery Nakariakov (1995--99), Robert Erd\'elyi (1996--98), Tony Arber (1996--00), Michael Ruderman (1997--00), Robert Walsh (1996--00), Dan Brown (1999--05), 
Thomas Wiegelmann (2000--02), Istv\'an Ballai (2000--02), Tibor T\"or\"ok (2003--04); also, some prominent research students here moved away without being postdocs, such as Moira Jardine (1985--88), Rekha Jain (1989--92), Erwin Verwichte (1996--99), Gordon Petrie (1997--2000), Danielle Bewsher (1999--2002), David Pontin (2001--04) and Antonia Wilmot-Smith (2004--07).

\subsection{St Andrews and the UK}

St Andrews has been a marvellous place to spend my research career. The beauty and history of the place and its location by the sea are very relaxing and condusive to pondering.  The university is of extremely high quality and is small enough that it is easy to interact with people from other disciplines and feel part of a supportive community.  Its smallness has also meant that we have come to know many students personally and to invite them into our home. The university hierarchy has always been responsive to ideas and easy to contact.  For example, it was a delight when the Principal agreed to my suggestion to rejuvenate the astronomy department by appointing Keith Horne, Andrew Cameron and Moira Jardine, and the university has also welcomed our suggestions to offer permanent positions to the AFs and URFs in the solar group.  In the life of the town outside the university, I have made good friends in several different communities (church, music and bridge). It is always intriguing to find new connections amongst people you know and to greet old friends returning to the town.  Increasingly we are meeting children of former students: for example, one the other day said ``Thank you for inviting me round for lunch on Sunday -- my parents say you did the same for them 25 years ago!"

Promotion and recognition is something I have never sought.  When Newby Curle (the head of department) came into my office  in 1977 and said  ``Congratulations, you have just been promoted to Reader from Lecturer" at the age of 33, I was over the moon (delighted) but staggered, since I had no idea I was being considered for promotion and had not thought it was anywhere near.  Six years later, exactly the same thing happened with promotion to full professor.  In those days, you did not apply for promotion yourself.

In the UK, I started as an applied mathematician and so was very much part of the Applied Maths community, attending their annual meeting regularly for a few years. Later, I was appointed on three separate occasions to the national Research Assessment Exercise Panel for the subject, which had the huge responsibility of assessing the research of every applied mathematician in the UK.  It started out as a relatively light-weight enterprise, but later became much more bureaucratic and burdensome, without any gain that I could see in depth or accuracy of assessment.

However, I also became more of a solar physicist and so naturally became interested in the UK's Royal Astronomical Society, of which the solar physics community (which has its own national meeting) is a part. I very much feel that it is important not to become too insular or isolated as a community and for solar physics to strengthen links in two directions.  The first is with astronomy as a whole, so as to communicate the vitality and important developments in solar physics, which have many potential implications elsewhere in the cosmos, while at the same time learning from astronomy how fundamental plasma processes (waves, instabilities, reconnection, turbulence, coronal heating, winds, jets and particle acceleration) are operating under different parameter regimes from the Sun.  The second direction is with solar system physics, especially magnetospheric physics, so that we can understand the influences of our Sun on space weather and climate.  Indeed, 25 years ago, Dave Southwood and I decided to set up a series of annual introductory and advanced summer schools on solar system plasmas which continue to this day and which are attended by all UK research students in solar and magnetospheric physics. I wanted to give a good grounding to the students, to open their eyes outside their own research topic and give them a wider appreciation, and also to help them feel part of a wider community, so that they could forge research links in future.   Developing links with astronomy and  the solar system becomes a heavy burden if it relies on a few people, but it is much easier if everyone plays a part.

\section{America Here I Come}\label{sec5}

\subsection{High Altitude Observatory (HAO)}
In 1971, we went to HAO in Boulder for a year -- then situated on the university campus.  It was quite an adventure, travelling by way of Iceland (for a few days, where you would go indoors to a much {\it hotter} environment heated thermally from hot springs) and New Orleans (where Clare's brother is a jazz trumpeter and where you would go indoors to a much {\it colder} environment due to air conditioning).   Every day I would come home from work with new words that I realised had a different meaning in the American language -- one day, when someone at a bus stop asked us ``how are you {\it guys} today?", Clare was quite shocked and commented ``Can't he see that I am a woman?"  

Being at HAO completely changed the way I view research.  I was thrilled by the openness and vitality and  the new concepts.  All the staff had their doors open, the seminars were inspirational and the staff were only too willing to describe and discuss their ideas and to collaborate. This was an exciting environment where communication and collaboration were key.  I began papers with: Marty Altschuler on the effect of Hall currents in a generalised Ohm's law in the corona; Yoshi Nakagawa on granulation power spectra and magnetoacoustic-gravity waves; Dean Smith on current sheets in coronal streamers and current limitation in flares; and Jerry Pneuman on solar angular momentum loss.   

I was filled with a desire to try and build a group back in St Andrews that had the best of what Cowling had taught me (high standards, in-depth research, critical but positive approach) and what I had found at HAO (vitality, openness to new ideas, collaboration, and above all being inspired by the latest observations).  Listening to the observers I met in Boulder was a real revelation. It was so interesting to hear about all the diverse phenomena in the solar atmosphere at first hand and to try and use my physical understanding of magnetic field behaviour to come up with (partial) explanations for what was being seen. High-quality observers such as Jack Harvey, Dick Canfield, Dave Rust, Hal Zirin seemed to know so much more than you can read in books, so my advice to a budding young theorist would be to pick a knowledgeable observer and listen carefully to what he or she can tell you.  

Although I did not like some of the superficiality I found in the USA nor the extreme evangelical right-wing politics, I very much appreciated the kindness, openness and generosity of so many Americans whom I  met.  They were less tied to possessions than in the UK and were much more willing to lend them. We returned home having our eyes opened to many good new patterns of behaviour, such as being friendly to newcomers and inviting them round to meals, or hosting ``pot-luck" suppers.

\subsection{Summer Visits}
In the next few years we returned to Boulder several times for summer visits, where collaborations continued and blossomed. For several summers we also visited France  (one in Meudon and one in Nice), Switzerland (swapping houses with Arnold Benz) and Germany (Munich), but later developed a pattern of spending a month or two in the USA each summer. When the children started coming along, the summer visits became great holidays for them and I would usually work from 6 am through to 2 or 3 pm and then join the family for fun. Often, we would swap houses and cars with someone, which would be highly convenient financially and provide us with a home environment to stay; they have always worked out really well.  Invariably, I would link up with someone with whom I resonated research-wise, but would not work out a detailed plan in advance. Instead, on my arrival, we would have an open discussion of ideas and would toss ideas around, invariably focussing on some new unexpected topic and working it out during the rest of the visit. 

\begin{figure}
   \centering
   \includegraphics[width = 9cm]{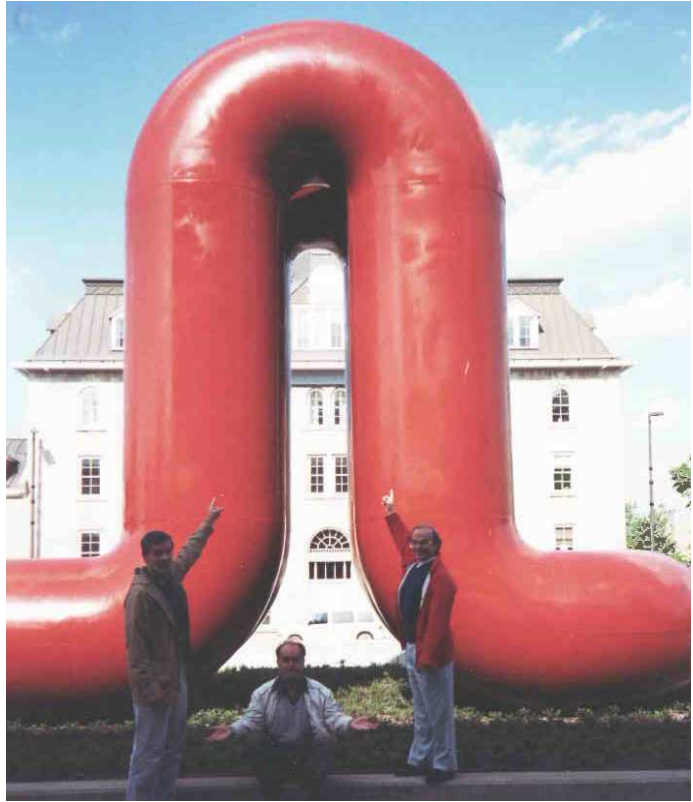}
   \caption{Dick Canfield, John Brown and I gaining inspiration about emerging flux from a piece of modern art in Ottawa, 1993.}\label{fig4}
\end{figure}

In 1973 in Boulder, I bumped into Dick Canfield in the mail room at HAO and asked him what was new at Sac Peak -- he replied ``Dave Rust has discovered distant H$\alpha$ brightenings  at the start of a flare with new flux emerging somewhere between them" (Figure \ref{fig4}).  Immediately, I thought of reconnection between the new and overlying magnetic fields, and so the three of us agreed to develop the idea, while I enlisted the help of Jean Heyvaerts to work out turbulence details in a current sheet between the new and overlying flux. The basic idea was that a small flare would be triggered when the current sheet reaches a critical height for turbulence onset and that if the overlying field contained a lot of stored energy it could be released as a large flare.  We wrote the basic ideas up in a couple of papers with authors listed alphabetically \cite{canfield75,heyvaerts77}, while Jean and I developed some of the details separately \cite{priest74b,heyvaerts76}. 

In 1974, we had a summer visit to Dartmouth College, New Hampshire, to work with Bengt Sonnerup, a wonderful researcher whom I admire greatly.  We started by looking at a paper by Gene Parker (another of my heroes, whose papers are always so stimulating) on a kinematic model for the transport of a unidirectional field by a stagnation-point flow.  One morning, I went into Bengt's office enthusiastically saying that I had just discovered that the solution was not just kinematic (satisfying the induction equation), but that it also satisfied the other  MHD equations including the equation of motion and so was an exact solution of the full nonlinear equations. He opened his drawer and pulled out a piece of paper with the same reasoning, so we had each made the discovery at the same time \cite{sonnerup75}.

\begin{figure}
   \centering
   \includegraphics[width = 10cm]{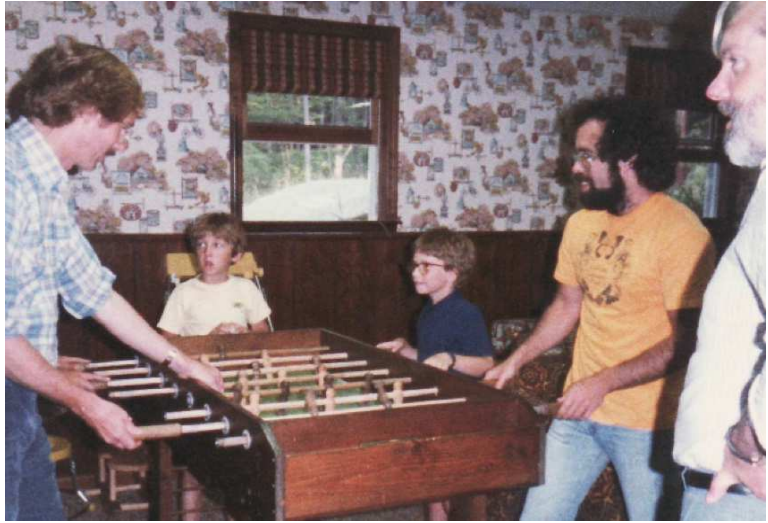}
   \caption{Marty Lee battling Andrew Priest and Phil Isenberg, refereed by Justin and Len Fisk, June 1983.}\label{fig5}
\end{figure}
Later, we made many summer visits to Durham, New Hampshire (Figure \ref{fig5}), to work with Terry Forbes (Section \ref{sec7}), where we continued our long-term collaboration begun when he came to St Andrews as a postdoc (1980--84). Durham was a marvellous environment with a great group led by Len Fisk and including some real characters who are highly talented, including Joe Hollweg and Marty Lee.  We would enjoy trips out to swim in a lake or explore the Maine coastline or even further afield to spend a long weekend in gorgeous Vermont with Rich Wolfson or to a farmhouse with no electricity belonging to Marty's sister, where the children enjoyed milking goats.  

We also spent one summer in Fairbanks Alaska, working with Lou Lee and exploring the amazing Denali National Park and the Columbia Glacier. With the sun hardly setting at all it was strange to see children playing outside until midnight, and an amusing incident occurred on midsummer's day when the institute had a great party with people dressed up in pioneer costumes from 100 years ago. One of the main events was a beard competition, and since beards are such sensitive parts of a man's anatomy it was inappropriate for the judging to be done by a local, so they asked my wife Clare to be the judge.  Well, it was easy enough to judge the longest beard or the bushiest beard, but when it came to the most handsome beard or the sexiest beard it was much trickier -- nevertheless, Clare did manage to carry it off with her usual tact and a sweet smile.

\begin{figure}
   \centering
   \includegraphics[width = 11cm]{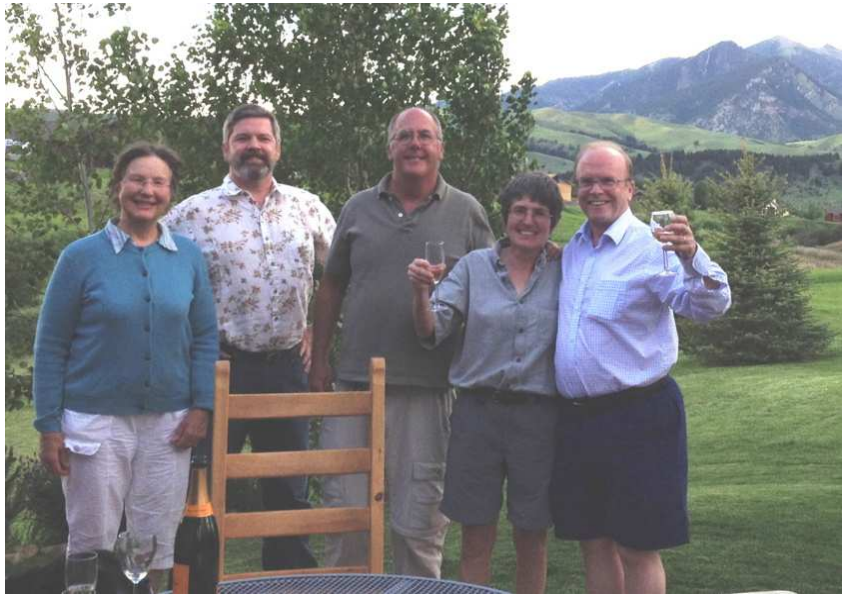}
   \caption{Summer 2013 in Bozeman with Clare, David McKenzie, Dana and Valerie Longcope.}\label{fig6}
\end{figure}
More recently, we have summered in Bozeman, Montana, as the guest of the fantastic solar group set up there by the inimitable Loren Acton, with a vibrant mixture between space instrumentation and observations, data analysis and theory, and now containing some outstanding scientists (Figure \ref{fig6}).  It was good to team up with two old friends, Dick Canfield and Piet Martens, and to meet and start collaboration with new ones, such as Dana Longcope.  It has also been marvellous to learn about \textit{Yohkoh} and \textit{Hinode} observations from David McKenzie and about interpreting coronal observations from Jiong Qiu. Each summer they run a superb 10-week programme for  third-year undergraduates, introducing them to the notion of research by giving them lectures (including four from me) and supervising a research project that is usually an extension to one of the ongoing group projects. Two or three students have come each year from St Andrews and have found it an inspirational experience, many of them subsequently continuing to a solar PhD.

\begin{figure}
   \centering
   \includegraphics[width = 12cm]{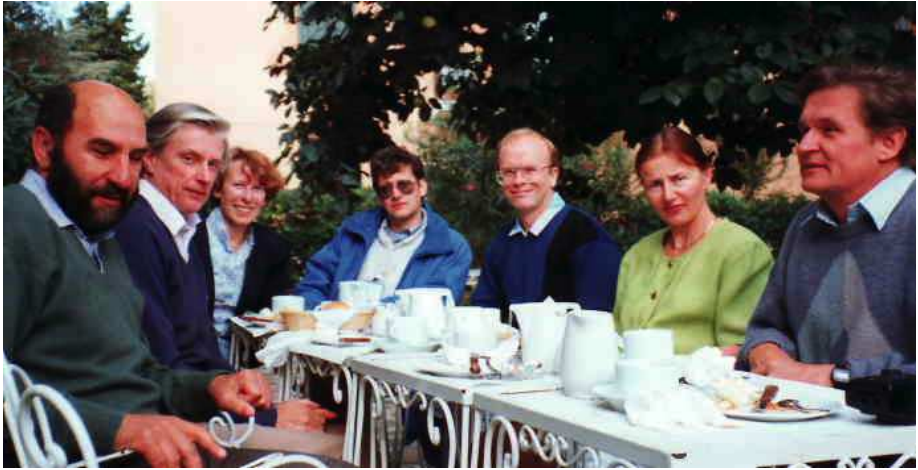}
   \caption{A snack at a workshop in Elba, September 1993, with stimulating company (Grisha Vekstein, George Simnett, Brigitte Schmieder, Aad van Ballegooijen, Galya Vekstein and Rainer Schwenn.}\label{fig8}
\end{figure}
\begin{figure}
   \centering
   \includegraphics[width = 8cm]{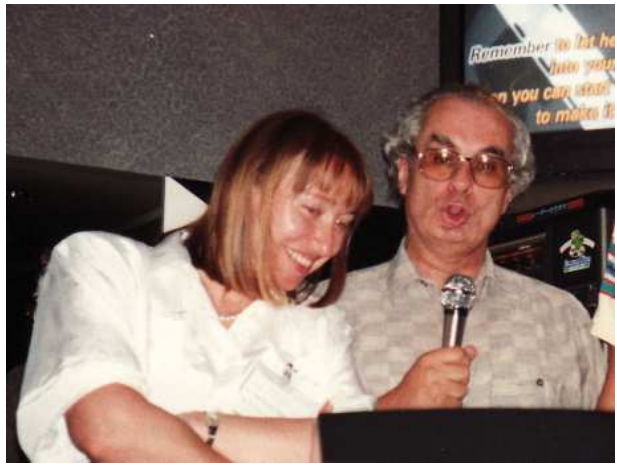}
   \caption{Lidia van Driel-Gesztelyi and Len Culhane practising their karaoke skills at the Kofu conference, Sept 1993.}\label{fig9}
\end{figure}
\begin{figure}
   \centering
   \includegraphics[width = 11cm]{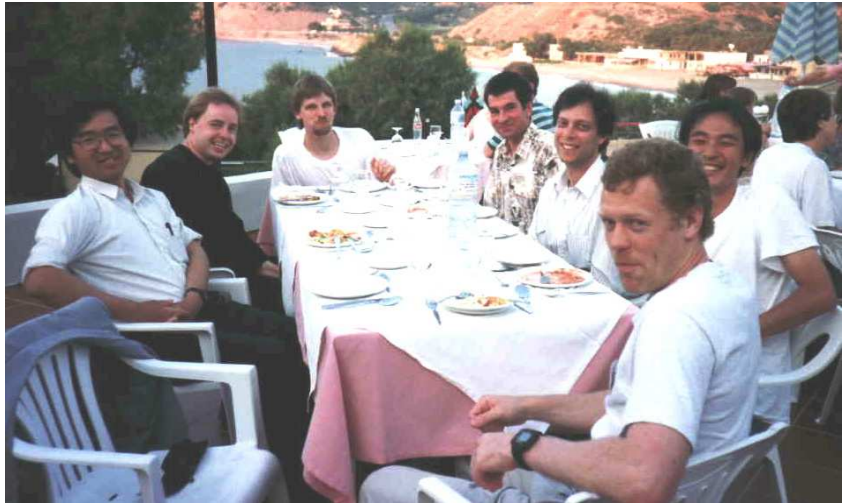}
   \caption{Relaxing by the Mediterranean in Crete, June 1995, with Kazunari Shibata, Bertil Dorch, Lutz Rastaetter, Gunnar Hornig,   Mitch Berger, Takaaki Yokoyama and Klaus Galsgaard.}\label{fig11}
\end{figure}
\begin{figure}
   \centering
   \includegraphics[width = 11cm]{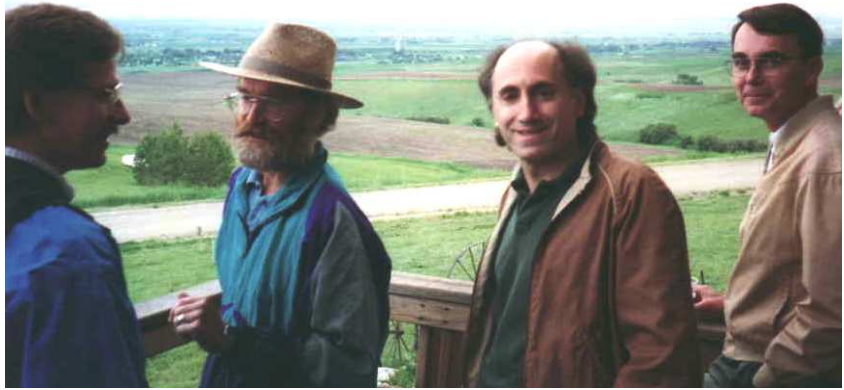}
   \caption{At Loren Acton's home at the AAS Solar Physics Division meeting in Bozeman, June 1997, with Jim Klimchuk, Ron Moore, Spiro Antiochos and Jack Harvey.}\label{fig12}
\end{figure}
\begin{figure}
   \centering
   \includegraphics[width = 11.8cm]{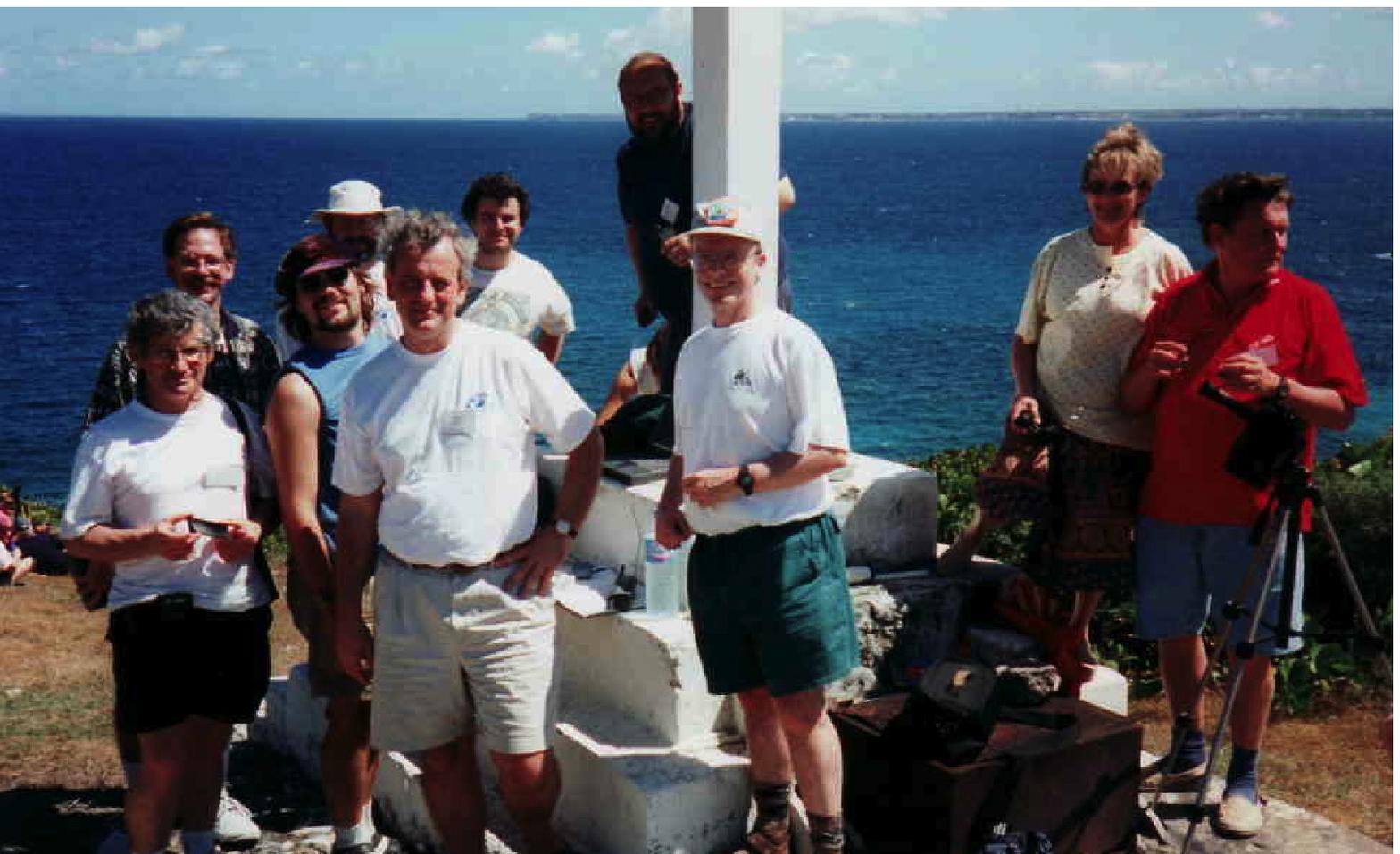}
   \caption{Viewing my first eclipse in Guadaloupe, February 1998, with Marcel Goossens, Jim Klimchuk, Craig DeForest, Bart De Pontieu, Robert Erd\'elyi and Rainer Schwenn.}\label{fig13}
\end{figure}
As well as stimulation from summer visits to the USA and different parts of Europe, there were many conferences and research visits further afield, including China (1982, Kunming), Japan (1982 Hinotori symposium), Russia (1985), India (1985 and 1989), Argentina (1991), Israel (1996), New Zealand (1997), Guadaloupe (1998, my first eclipse), Ecuador (2001),  and elsewhere. These really opened my eyes in the 1970s, 80s  and 90s, enabled me to make many new friends and helped me appreciate different cultures (Figures \ref{fig8}, \ref{fig9}, \ref{fig11}, \ref{fig12}, \ref{fig13}, \ref{fig7}).

\section{The Book ``Solar MHD"}\label{sec6}

In the mid-1970's I was visiting Meudon observatory and staying in the rooms there, when we had a late-night discussion with several other visitors over a drink.  The conversation turned to books and Josip Kleczek asked each of us which new book we would like to see written. I said I thought there was a need for a book on solar magnetohydrodynamic theory and mentioned that I would consider writing it after another 40 years, towards the end of my career when I knew more.  He replied: ``You should write it now, while you are enthusiastic and have lots of ideas", so with his encouragement I decided to do just that and after another 7 years the book {\it Solar MHD} was born \cite{priest82}. I only hope it has been of some use to young researchers wanting to enter the field. The aim was to summarise solar observations and the basic equations, before reviewing the different branches of MHD theory.

Of course, now most of the book is completely out of date, and so, for the past 10 years I have been writing from scratch a completely new replacement for  {\it Solar MHD}.  In November, I finally finished the page proofs and so hopefully, by the time you read this, it has appeared in spring 2014, published by Cambridge University Press \cite{priest14}. This book is a completely new rewrite, not just a new edition, so I had to decide on a new name for the new ``baby". In the end, I came up with {\it Magnetohydrodynamics of the Sun}, so as to indicate that the subject matter is the same as before, but the book is very different.  

In the new book, the basic equations and basic theory are the same, but the observations have been revolutionised by high-resolution observations from the ground and space, and so there is a new 80-page description of observed solar features.  Also, the basic theory has been developed greatly, and so there are new in-depth accounts of force-free equilibria and of waves in nonuniform medium and a new  70-page chapter on magnetic reconnection.  The applications of the basic theory have also been completely rewritten from scratch, including chapters on dynamos (with a comparison of flux-transport and tachocline dynamos) and magnetoconvection and sunspots. Atmospheric heating by waves or reconnection is summarised, and the new theories for prominences as sheets in three-dimensional flux ropes are reviewed. A chapter on the MHD of flares and CMEs describes the possible causes of an eruption and the nature of the energy release by reconnection. Finally,  wave-turbulence and reconnection models of the solar wind are summarised.

\section{Key Long-Term Collaborations}\label{sec7}

\begin{figure}
   \centering
   \includegraphics[width = 11.8cm]{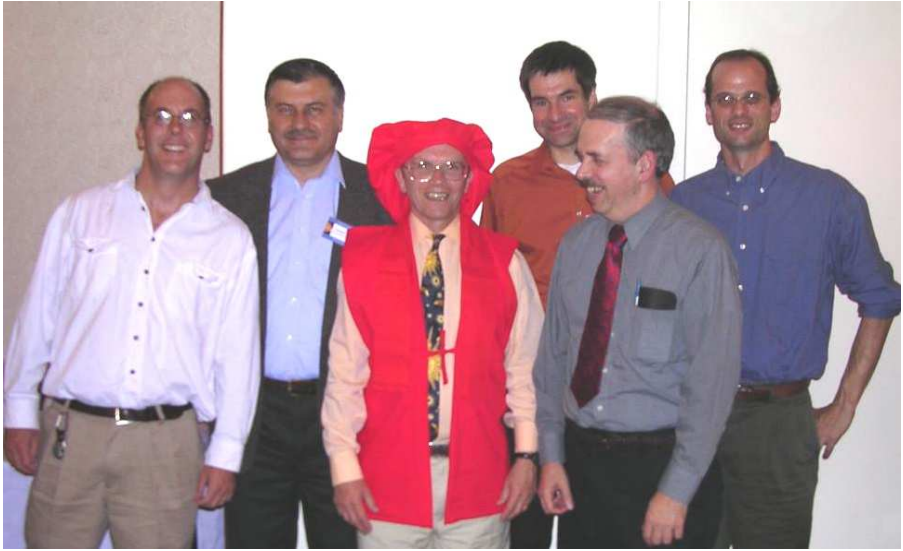}
   \caption{Some of my collaborators at the September 2003 Reconnection Conference in St Andrews (my 60th birthday): Dana, Slava, Gunnar, Terry and Mark Linton. I am wearing a traditional Japanese jacket and hat (aka-chan, symbolising the start of a new life of health and happiness) presented by Daikou Shiota, Ayumi Asai, Takehiro Miyagoshi, Noriyuki Narukage and Syuniti Tanuma.}\label{fig15}
\end{figure}
I have been fortunate to be close friends with several highly talented researchers, with whom I have collaborated extensively.  In each case, there has been an easy and stimulating communication, which has come from the facts that I like them very much, that we have much in common scientifically in terms of our approach, understanding and high standards, and that the differences in our backgrounds and skills bring new ideas.  Thus, I have basically a simple set of analytical skills and physical ideas and a wide range of interests, whereas Terry Forbes has numerical skills and likes probing to the core of a problem in an original way and Jean Heyvaerts had deep physical understanding and impressive mathematical skills that he brought to bear in a powerful way once he had the bit between his teeth on a problem.  Pascal D{\'e}moulin is a really fast worker who explores all the physical aspects of a new problem in a rapid sweep.  Gunnar Hornig is an incredibly deep thinker on all aspects of three-dimensional reconnection and tends to approach a topic from general to particular, whereas I do so in the opposite direction and come to an understanding much more slowly by trying to clarify various aspects of his thinking with simple examples.   Dana Longcope is one of the smartest MHD theorists in the world, with a much better understanding of statistical matters than me and an impressive way of working out ideas on a white board (Figure \ref{fig15}). 

\begin{figure}
   \centering
   \includegraphics[width = 11.8cm]{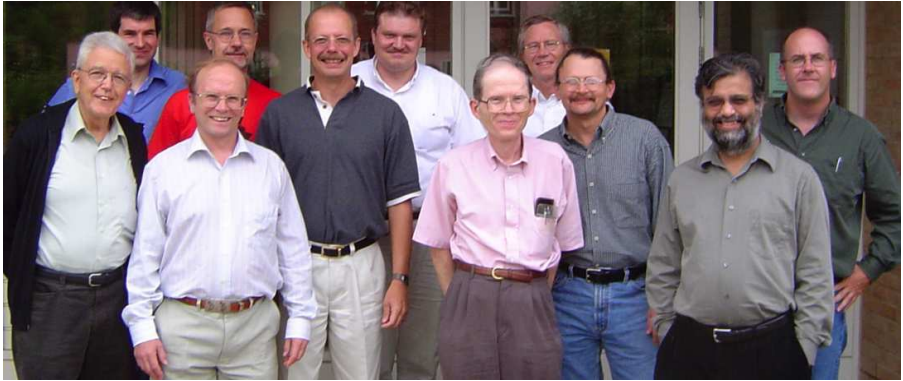}
   \caption{Some superb reconnection theorists at the first annual reconnection workshop in 2004 in Cambridge.}\label{fig16}
\end{figure}
Several of these collaborations have been helped by a series of Annual Reconnection Workshops, which Terry Forbes, Joachim Birn and I started at the Isaac Newton Institute in 2004 (Figure \ref{fig16}). They involve collisionless and MHD theorists from solar physics and space physics and alternate between Europe and the USA.

\subsection{Terry Forbes -- Fast Reconnection and Computational Experiments}\label{sec7.1}

I had never met Terry when I appointed him as a postdoc, though I had been impressed with his magnetospheric MHD papers, and so we arranged to meet at the airport in Edinburgh on his arrival in 1980.   In order to ``help us" recognise each other but unknown to me, a mutual friend Joachim Birn gave the following description of me to Terry: ``Eric is very distinctive -- a tall thin very old gentleman with ginger hair, who always wears a suit and tie".  When everyone from the flight had disappeared to leave just two of us in the arrivals area, Terry at last realised that Joachim had been joking and that the short young person in the Hawaiian shirt must be me.

\begin{figure}
   \centering
   \includegraphics[width = 10cm]{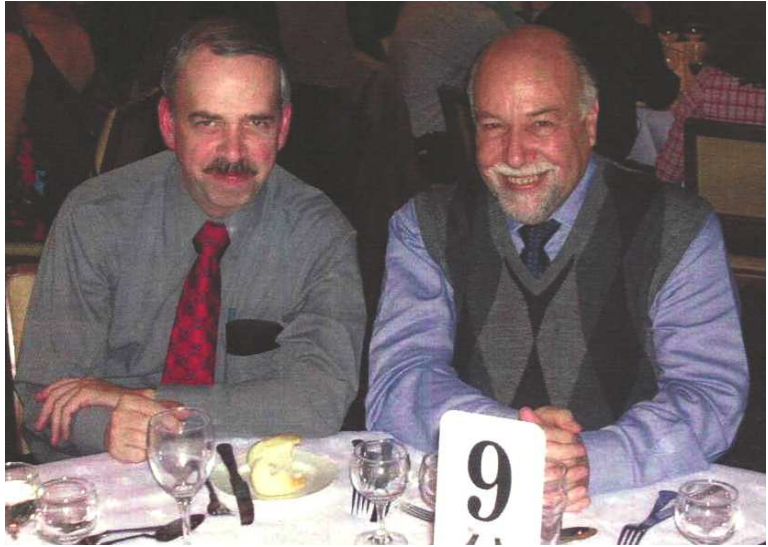}
   \caption{Two of my closest friends and collaborators, Terry Forbes and Jean Heyvaerts, at the conference dinner in honour of my 60th birthday, September, 2003, in St Andrews.}\label{fig17}
\end{figure}
Terry fitted in well at St Andrews and was a sociable friendly addition (Figure \ref{fig17}). He developed an expertise in computational MHD and, together with a couple of outstanding research students (Peter Cargill and Philippa Browning) and  Alan Hood (then working in Edinburgh), they formed a very lively group.  We soon got on like a house on fire and worked on a variety of solar flare topics (2D computational models).  The first was to study the way in which reconnection at the start of a flare is initiated by tearing and develops nonlinearly into Petschek reconnection in a current sheet that is stretched out below an erupting flux rope \cite{forbes82a,forbes83a}. A second topic was to investigate the complex evolution of a current sheet after the onset of an enhanced turbulent magnetic diffusivity \cite{forbes82e}. Thirdly, we decided to model  the effect of emergence of new magnetic flux from below the photosphere \cite{forbes84a}. We also discovered a way of calculating the electric field in the flare site from the motions of the flare ribbons and the field strength at the flare ribbons \cite{forbes84b}.

Our collaborations have continued ever since he moved to New Hampshire.  One highlight was our discovery of different regimes of fast reconnection \cite{priest86a}. Petschek's mechanism had been accepted for many years as the valid and standard model for fast reconnection (i.e., reconnection much faster than Sweet-Parker and therefore potentially capable of explaining flare energy release), although very few people had read the original paper in detail.  But then grave doubt on the mechanism was cast by Biskamp,  who had performed some impressive numerical experiments which produced much slower reconnection than Petschek.  He claimed that the results are completely independent of the boundary and initial conditions.  However, as an Applied Mathematician, I was suspicious, since I know how important boundary conditions are, so Terry and I decided to derive Petschek's mechanism in our own way and confined to a computational box rather than being in an infinite domain.  

We linearised the basic MHD equations about a uniform field in the upper half of the numerical box. We studied the nature of the inflow regime and joined it to the diffusion region and the regions between two pairs of standing slow-mode shocks.  We were delighted to find a range of Petschek-like solutions that depend on the size of the dimensionless flow speed at the inflow boundary ($M_e$, the reconnection rate). Whereas Petschek had given a rough physical argument for there being a maximum reconnection rate, we were able to vary a parameter in our solutions (the inflow speed at the diffusion region for given magnetic Reynolds number, $R_m$), and we were thrilled to find that the reconnection rate does indeed possess a maximum value. Furthermore, this maximum is typically 0.01 or 0.1 and has the same scaling with $R_m$ as Petschek had predicted.

Now, one of the key aspects of being a theorist is to try and put yourself into the mind of the person who produced the theory you  are examining, since often the way of deriving it is more intuitive and tortuous than the clean way in which it is written up. Also, what are the assumptions, especially those that are not stated?  And how can you change those assumptions in order to make the analysis more realistic or more relevant?

Well, in this case, we and Petschek had assumed the inflow region is current-free or potential and we had linearised about a uniform field in one inflow region and the oppositely directed uniform field in the other inflow region.  So, the first extension we considered was to include pressure gradients in the inflow region. This led to an extra constant ($b$) in the inflow solutions, which we could relate to the inclination of the flow velocity on the inflow boundary.  We had discovered a completely new family of {\it almost-uniform} fast reconnection solutions (with $b=0$ giving Petschek's mechanism).  Furthermore, we realised that the solutions are determined not only by the magnitude of the inflow on the boundary but by its inclination: thus, converging flows tend to compress the plasma in the inflow and to lead to small diffusion regions, whereas diverging flows expand the plasma and produce longer diffusion regions.
Later, we generalised the models in a different way by, as Biskamp had essentially done, linearising about an X-point field rather than a current sheet (with uniform field on either side) and this led to another new family of {\it nonuniform} fast reconnection solutions.  Later, we undertook an in-depth comparison of numerical and analytical models \cite{forbes87}, which gives a comprehensive understanding of Biskamp's solutions.  The bottom line is that the nature of the solutions depends on the initial state and the boundary conditions and therefore on the particular application you have in mind. 

Another highlight, building on earlier work by Piet Martens, was the discovery that a flux rope can evolve through a series of equilibria and erupt at a catastrophe when the equilibria ceases to exist \cite{priest90a,forbes94}.  In addition, we coauthored a book that reviews magnetic reconnection theory and its applications \cite{priest00}.  More recently, we have been helping Hubert Baty to explore the effect of nonuniform magnetic diffusivity on fast reconnection \cite{baty06,baty09a,baty09b}.

\begin{figure}
   \centering
   \includegraphics[width = 11cm]{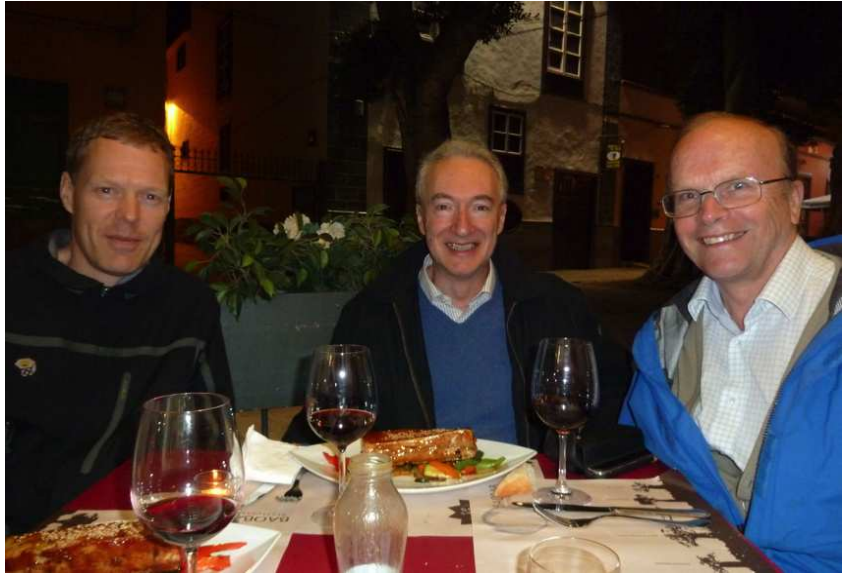}
   \caption{Dinner outside (January, 2014) in Tenerife with two good friends, Klaus Galsgaard and Fernando Moreno Insertis.}\label{fig18}
\end{figure}
Terry's long-lasting influence on the group was to show us the importance of computational experiments, which have since been continued as a key ingredient of the group's work by Alan Hood, Clare Parnell, Ineke De Moortel, Duncan Mackay and Vasilis Archontis, with the continual help of the incomparable Klaus Galsgaard (Figure \ref{fig18}). Computational MHD modelling was very much a new tool in those days, with several pros and cons.  On the negative side, you can only model a few values of the dimensionless parameters such as the magnetic Reynolds number ($R_m$) or the plasma beta ($\beta$), and in the case of $R_m$ the values you can model are far from realistic.  However, on the positive side, computational models can reveal new complex behaviour that is often not predicted analytically, and so there is an exciting sense of discovery in these ``experiments".  It remains the case that the most effective approach is often a complementary one of computational modelling, physical insight and analytical modelling.  Indeed, if the computational experiments are not analysed in depth with physical and analytical insight, the results are often superficial and fleeting in their impact.

Terry is a joy to work with and wonderful company.  He remembers amusing events much better than me and is a brilliant raconteur.  He is also a skilful speaker, usually focussing on one or two key points and describing them with great clarity, and never falling into the common trap of putting too much into a talk.

\subsection{Jean Heyvaerts -- Emerging Flux, Coronal Heating by Phase Mixing, Turbulent Relaxation or Flux-Tube Tectonics, and the Importance of Magnetic Helicity in the Corona}\label{sec7.2}

In the early 1970's I met Jean Heyvaerts at a conference and he persuaded me to attend a small workshop in Aussois organised by Andr{\'e} Mangeney. I arrived and, looking around at the dozen or so other people at the workshop, I realised that I was the only non-French person present, so naturally we were going to spend the week discussing in French. I had done French at school for a few years, and, although I had learnt to read and write reasonably, there had been virtually no spoken 
French.  Immersing myself into the music of the language in Aussois was amazing, but in the evening I would lie down in my bed with the sounds of countless new words echoing around in my brain like a hornet's next.  It was a real workshop with simply a free-floating discussion and nothing planned.  Andr{\'e} had brought along a series of papers on astrophysical MHD and so we picked out the few that interested us, read them and described them to the others, leading to a discussion of new directions for study.

I have since then always enjoyed visiting France and speaking French, which certainly helps understand  a little better the people and culture, which I like very much. Since I had done Latin at school, I could usually estimate whether an English word has a root related to French or not, and so did a lot of guessing -- speaking the English word with a French accent and hoping for the best (and trying to avoid ``faux amis"). I can remember on one occasion turning up in Nice to examine a thesis. Walking into the examination room, I detected a puzzlingly cool atmosphere from the audience -- until, that is, I said in French ``I presume we are going to conduct this examination in French", when the worried expressions turned into smiles. I also remember the delight at the end of the talk, after {\it le jury} had pronounced our (positive) verdict, when the lecture doors opened to reveal the banquet with delicious trays of food and barrels of wine.

Once, when I arrived at the gate at Meudon and told the guard my name (Priest), he looked down the list on his desk and said my name was not on it.  However, I looked at the list and pointed to it, at  which he said ``Ah, monsieur priiiii-est".  On another occasion, I had forgotten to take my dictionary to a local restaurant and noticed a series of dishes ending in ``de veau", and thought it would be good to have a tender piece of veal in a tasty sauce. I just  chose by random in the list something called ``cerveau de veau", but was rather surprised when later a plate arrived with a huge veal brain like a massive half cauliflower -- not exactly what I expected.

I only wish  I had learned several other languages as a school or university student.  For 6 months before visiting a new country,  I used to try to learn a smattering of the language with tapes and books, practising every day. However, the problem was that in each case, whether it be Spanish, German, Italian or Russian, my brain was too old and, after returning from the visit, my brain rapidly emptied itself of what I had tried to learn.

\begin{figure}
   \centering
   \includegraphics[width = 11.7cm]{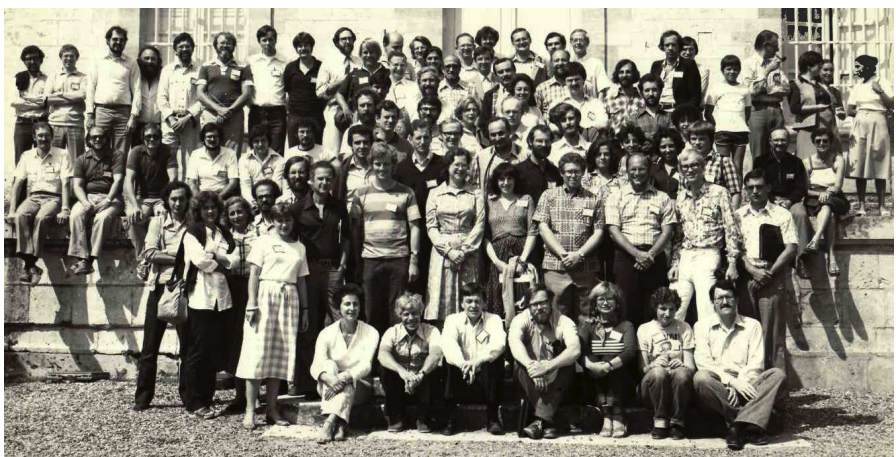}
   \caption{The Bonas conference in September, 1980, in which you can see many distinguished members of our community, looking a little younger than now, including on the front row Roger Bonnet, Ian Roxburgh and Andrea Dupree, on the second row Randy Levine, Claudio Chiuderi, Alan Hood, Carol Jordan and Bob Noyes, on the third row Henck Spruit,  Georgio Einaudi, Piet Martens, myself and Gene Parker, and on the back row Andy Skumanich, Peter Ulmschneider, Jean Heyvaerts, a hairy Steve Schwartz and Kes Zwaan (holding a pipe).}\label{fig7}
\end{figure}
Our first papers were in the mid-70's on the emerging flux model for flares, culminating in Heyvaerts, Priest and Rust (1977), in which Jean's main contribution was to work out the conditions for onset of plasma turbulence in the current sheet  (\textit{c.f.} Section 5.2).  We both had great fun at a ground-breaking conference on the solar-stellar connection organised by the charismatic Roger Bonnet in Bonas in 1980 (Figure \ref{fig7}). This was the first time my mind had been opened properly to the diversity of solar-like behaviour on other stars.  A memorable event was Andrea Dupree dancing on the table with Jeff Linsky in the evening after a visit to some armagnac caves;  many were the worse for wear the following morning.

Jean would visit St Andrews for a month or I would join him in Paris, Nice or Strasbourg, and we would generally have no particular plan in mind, but would just share the ideas buzzing around inside our heads. Suddenly, we would  just focus on one particular idea which grew from nothing and demanded our attention with a momentum of its own. Then we would use all the skills and techniques we had honed over many years to help it come to fruition.  Jean would work at a feverish pace and give me each morning pages of dense working to check over and comment on, and usually after a month of intense activity the idea would be written up and submitted.  

One example of this was the paper on coronal heating by phase mixing, where together we realised that there was an elegant way of treating standing or propagating waves in a simple unidirectional magnetic field whose strength varies with position \cite{heyvaerts83}.  We derived the basic equation fairly quickly, but then had to wrestle with it to find ways of treating it analytically and also to estimate the importance of tearing-mode or Kelvin-Helmholtz instabilities in the phase-mixed waves.

Another example was the idea of using Taylor relaxation to heat  the corona by MHD turbulent reconnection \cite{heyvaerts84}.  I had read Taylor's papers on turbulent relaxation in a laboratory machine as a means of explaining the linear force-free nature of the field, and mentioned it to Jean in St Andrews. I thought that perhaps the corona too was turbulent and, indeed,  we were the first to suggest that magnetic helicity could well be an important invariant in the solar corona. Perhaps, we thought, it was gradually built up by photospheric motions and ejected during prominence eruptions when it became too great.  Jean immediately agreed this was a great idea, but how to work it out mathematically? -- given that Taylor's analysis was for a closed volume, whereas we wanted to have the field entering and leaving the coronal volume.  This in turn meant that we had to deal with the tricky issue of gauge invariance of the magnetic helicity.  We worked out a technique to do so and applied it to coronal arcades and coronal loops.  Later, we found out that independently, Mitch Berger had  come up with a brilliant idea of relative magnetic helicity, which was equivalent to our technique but more elegant and so we immediately appreciated his approach.

Later, the amazing Philippa Browning applied these Taylor relaxation ideas to coronal arcades and collections of flux tubes \cite{browning86b,browning86c}, and Jean I applied them to an accretion disc \cite{heyvaerts89}.   Subsequently, we returned to the problem and substantially improved it by producing a theory for turbulent heating that is self-consistent, with the turbulent diffusivity and viscosity determined from  a cascade theory of turbulence \cite{heyvaerts92}.  This applied both to heating by waves and to heating by reconnection and has potential for future development and application.  The theory was in turn applied to a self-consistent turbulent accretion disc \cite{heyvaerts96a}.

\begin{figure}
   \centering
   \includegraphics[width = 11cm]{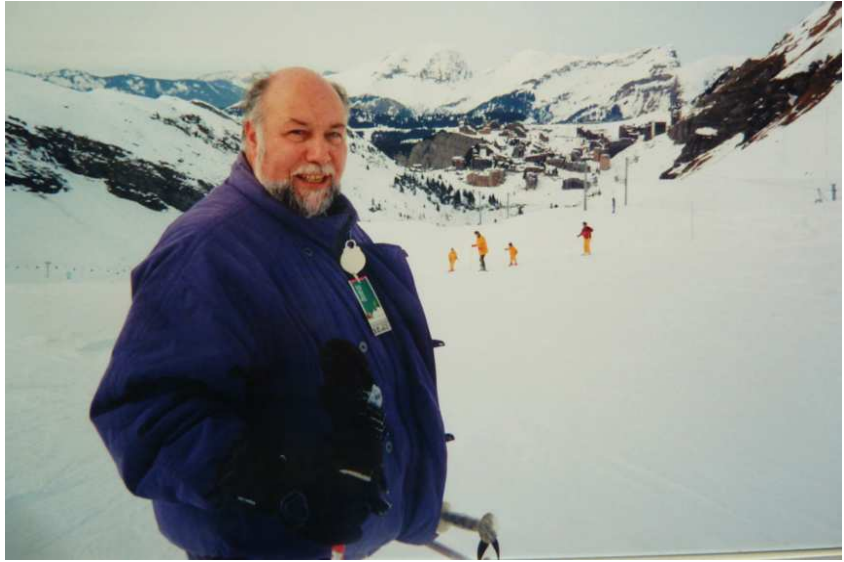}
   \caption{Jean Heyvaerts on the ski slopes with me in February, 1997.}\label{fig19a}
\end{figure}
On another visit to St Andrews, Jean and I decided to look in a new direction on  the coronal heating problem and wonder if the nature of the heating could be determined from observations of the temperature and density structure along a coronal loop.  We realised that, for very long loops, the temperature structure would be observably different if the heating were uniform, concentrated at the feet or concentrated at the summit of the loop, and that these would correspond to different heating mechanisms. A preliminary application of the idea to large Yohkoh loops suggested that in the coronal part of the loop above the footpoints the heating is rather uniform \cite{priest98a,priest00a}.  However, it is the idea rather than the preliminary result that is important, and it could certainly be applied fruitfully in future.

Later, after inspirational discussions with Alan Title (who is incredibly smart and interesting to swap ideas with), we came up with a ``Flux Tube Tectonics Model" for coronal heating \cite{priest02a,close04b}  (see also D{\'e}moulin and Priest, 1997).  This was essentially a development of Parker's classical braiding model in which we included the {\it magnetic carpet}, the fact that coronal magnetic field does not link to the surface in regions of almost-uniform flux (as Parker had assumed), but in highly concentrated intense flux tubes.  This makes a huge difference to the efficiency of producing current sheets, since Parker needed complex braiding of footprints, but in our case any relative motion of the flux tubes produces current sheets along the separatrices and quasi-separatrices that separate the distinct flux tubes from one another.  Of course, once we had come up with the physical idea, the problem was to produce an elegant way of demonstrating the formation of current sheets and estimating the energy flux in both two-dimensional and three-dimensional arrays of flux tubes.

Jean was such a pleasure to spend time with, over some wonderful food, discussing so many issues until the early hours and going skiing near Geneva for a long-weekend (Figure \ref{fig19a}).  Research-wise, it was also a joy to share ideas and to work out new theories.  I would know better the previous solar observations and theories, but he was a fantastic theoretical physicist with a brain that could operate in a tremendously powerful and original way.  I was desperately sad when he died last year and miss him very much.  

\begin{figure}
   \centering
   \includegraphics[width = 11.5cm]{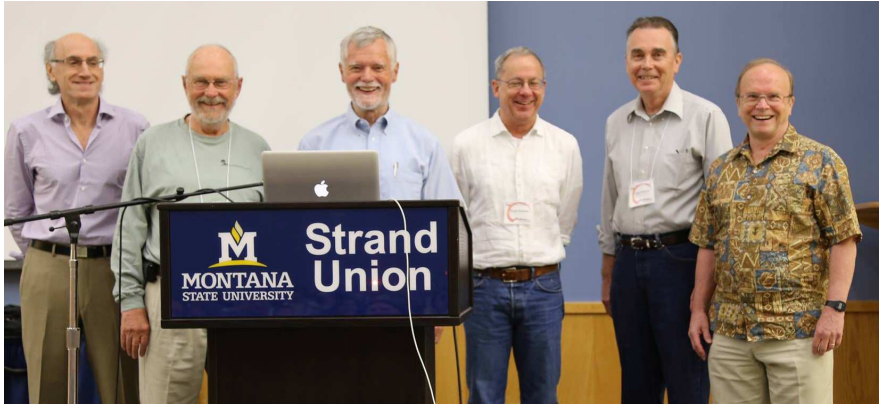}
   \caption{Hale prize winners at the Solar Physics Division meeting in Bozeman, June 2013. From the shirts, guess who is the non-American! Each of them have been inspiring examples to me of high-quality solar physicists and delightful people: Spiro Antiochos, Loren Acton, Dick Canfield, Hugh Hudson and Jack Harvey.}\label{fig14}
\end{figure}
\begin{figure}
   \centering
   \includegraphics[width = 11cm]{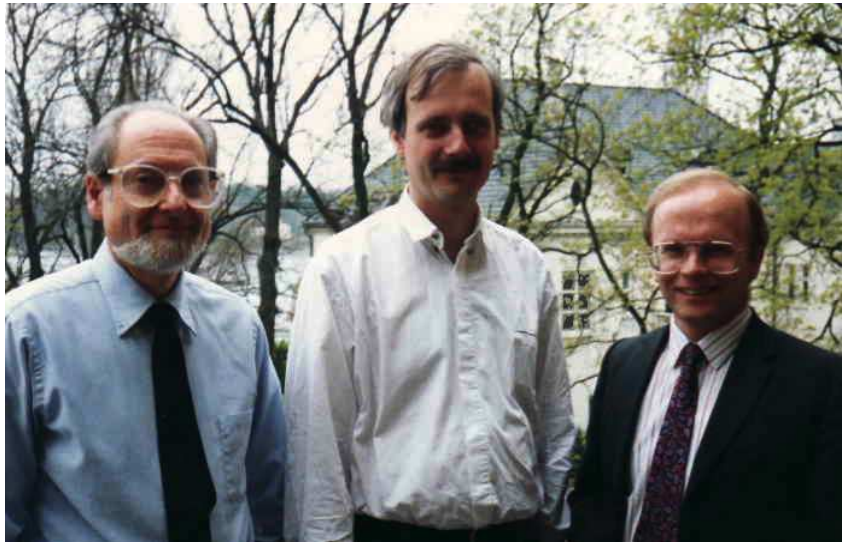}
   \caption{With Bob Stein and Mats Carlsson on the day we were all admitted to the Norwegian Academy, May 1994.}\label{fig10}
\end{figure}
It only makes me realise how precious life is and how we must treasure and appreciate every day, with its little moments of joy as we interact with someone or witness an act of kindness or have our eyes opened to the beauty around us or can help value someone through a kind word.  These are much more important than power or honours.  I have been delighted and humbled by several honours that have come out of the blue (such as Hale prize (Figure \ref{fig14}), membership of the Norwegian Academy of Sciences, Fellowship of the Royal Society (Figure \ref{fig10}), and this year an Honorary degree from St Andrews), but friends and individual people are much more important.

\subsection{Pascal D{\'e}moulin and QSLs}\label{sec7.3}

\begin{figure}
   \centering
   \includegraphics[width = 11cm]{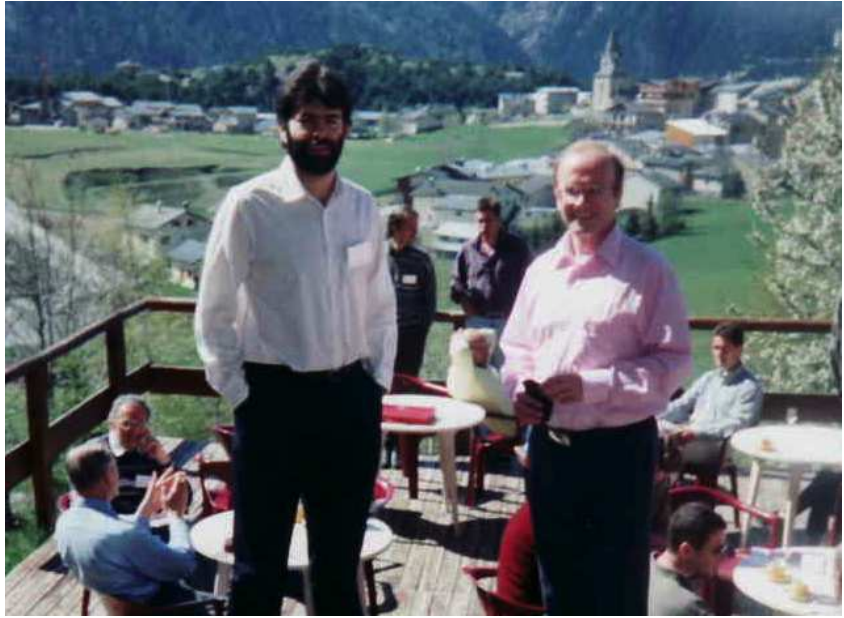}
   \caption{With Pascal at the IAU Colloquium in Aussois, April 1997.}\label{fig19}
\end{figure}

The present excellent solar MHD group in Meudon was born from the decision of Jean Heyvaerts to send a bright young research student (Pascal D{\'e}moulin) to work with Brigitte Schmieder and to suggest he go to St Andrews to learn some MHD in 1987--8.  It was a real pleasure to work with him, since he is so bright and couples a superb physical  understanding with great technical skills (Figure  \ref{fig19}).  He visited several times and we became good friends, working to begin with on the general topic of prominence equilibria, the formation of dips and their support and loss of equilibrium or instability in a force-free field \cite{demoulin88,demoulin89a,demoulin89b,demoulin89c,demoulin91b,demoulin93a}.

Since we were becoming accustomed to playing around with force-free fields, one day I wondered how the usual potential field solutions for point sources (such as monopoles and dipoles) would be generalised in a linear force-free field, and so suggested that Pascal take a look at it.  He tackled it with his usual energy and speed and we were surprised at how complex and different the solutions are \cite{demoulin92c}.  These solutions were later used to study the magnetic topology of active regions, by developing the work of Gorbachev and Somov (1988).  

Next, we worked out together a major new concept of three-dimensional reconnection at {\it quasi-separatrices (QSLs)} in the absence of null points \cite{priest95a,demoulin96b}, which Pascal has since applied with Cristina Mandrini and others to flaring active regions and has found that some flares occur at separatrices (across which the mapping of field lines is discontinuous) and others at QSLs (across which the gradient of the mapping function is large but not infinite).  Since I enjoyed our conversations so much,  I have from time to time wished we had continued collaborating, but instead Pascal's collaboration with Cristina has blossomed and led him to Argentina for visits, where they have made major contributions to understanding flares. 

Our development of the idea of QSLs had not just appeared from nowhere but was built on previous work of others and it was also an interesting resonance between our different complementary interests, as described below.   Schindler et al (1988), Birn et al (1989),  Hesse and Birn (1990) had proposed the revolutionary idea that, in general, 3D reconnection can (unlike 2D reconnection) occur in the absence of null points and separatrices and that it is associated with a component of electric field ($E_\parallel$) along the magnetic field. Such a component is associated with a non-ideal term in Ohm's law, since the ideal Ohm's law ($\bf E+\bv \x \bB=0$) implies that the electric field is normal to the magnetic field. 

On the one hand, Pascal had been interested in understanding the topology of active regions and so had developed techniques for modelling the photospheric field as a series of many sources and for calculating the positions of nulls, separatrices and separators for the resulting potential or force-free coronal magnetic field. Together with the Argentine group, this had been applied to several flares showing that energy release occurs near the computed separatrices  \cite{mandrini91,mandrini93,demoulin94a,demoulin94b}. However, a limitation of such source modelling is the need to integrate field lines below the photosphere, bearing in mind that a flare does not significantly affect the photospheric  flux distribution.  So what was the impact of sub-photospheric magnetic nulls on coronal physics, and how can you define coronal magnetic topology with sub-photospheric nulls?

On the other hand, Terry and I had been interested in the basic process of reconnection without a null point. We had first suggested that reconnection occurs at a {\it singular line}, along which $E_\parallel \neq 0$ and near which the field in a plane perpendicular to the field line has an X-type topology -- i.e., reconnection is associated with singularities in electric field in an ideal approach, which are resolved by diffusion \cite{priest89a}.  Also, we had introduced a germ of the idea of QSLs (but had  called them {\it magnetic flipping layers}) by studying a simple model field with $B_x=x,\ B_y=y,\ B_z=constant$ and had solved $\curl \bE=\bf 0$ and $\bE=\bv \x \bB=\bf 0$ in the ideal region, as well as calculated the mapping of field lines.

Then Pascal and I came together and produced three papers virtually simultaneously. The first paper introduced the term QSL, defined it as region with a steep gradient in the mapping function, and described the basic theory  \cite{priest95a}.  The second paper developed the theory further and applied it to an example with four photospheric sources \cite{demoulin96b}, while the third considered a twisted flux tube characteristic of the field around a prominence and calculated the thicknesses of the QSLs \cite{demoulin96a}.  There was also an important application to coronal heating \cite{demoulin97b}, in which we showed how current sheets can form along QSLs between many tiny flux tubes in response to generic photospheric flows, thus predating and anticipating many aspects of the coronal tectonics model.

\subsection{Gunnar Hornig -- 3D Reconnection and the Dundee Group}\label{sec7.4}

\begin{figure}
   \centering
   \includegraphics[width = 9cm]{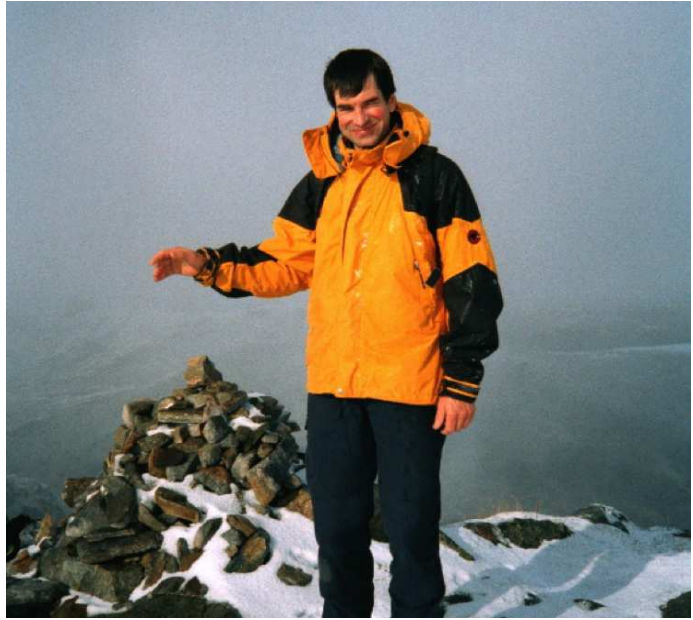}
   \caption{Walking in the Scottish hills in October 2002 with Gunnar Hornig.}\label{fig20}
\end{figure}

Gunnar  (Figure  \ref{fig20}) has had a huge influence on the St Andrews group and on the research done by a series of students and postdocs (Section \ref{sec8}).  He has taught me a lot of what I know about 3D reconnection and has a really deep understanding of it, of the nature of magnetic diffusion and of the differences between field line and flux conservation.  Now that my latest ``baby" has been born (the successor to {\it Solar MHD}), I hope to have time to spend discussing lots of questions that still remain in my mind.  In addition, he is the inspiration behind the impressive new group working on reconnection at Dundee University. 

We wrote together an account that stresses how completely different 3D reconnection at a finite diffusion region is from 2D reconnection at an X-point \cite{priest03}.  In two dimensions, a magnetic flux velocity can always be defined: two flux tubes can break at a single point and rejoin to form two new flux tubes. In three dimensions, on the other hand, a flux tube velocity does not generally exist. The magnetic field lines continually
change their connections throughout the diffusion region rather than just at one point. The effect of reconnection on two flux tubes is generally to split them into four flux tubes rather than to rejoin them perfectly. During the process of reconnection, each of the four parts flips rapidly in a virtual flow that differs from the plasma velocity in the ideal region beyond the diffusion region.

Around the same time, we developed an elegant technique for obtaining analytical models for steady, kinematic reconnection with an isolated diffusion region containing no null point \cite{hornig03}. In other words, we solved the steady equations $\bE+\bv \x \bB=\eta \curl \bB, \curl \bE=0, \div \bB=0$ and $\div(\rho \bv)=0$, starting with a basic magnetic field of the form ($y,kx,1$) with a uniform current.  The resulting solutions show how rotational plasma flows are produced above and below the diffusion region, which are related to the small change in magnetic helicity that is always associated with 3D reconnection. 

David Pontin and Antonia Wilmot-Smith were two wonderfully gifted research students of mine, who continued this work by applying a similar technique in various ways with Gunnar's help, and they are now tenured members of the Dundee group.  David focused on reconnection at null points. He modelled the effect of a spine-current at a spiral null with a field ($x-jy,y+jx,-2z$) \cite{pontin04a} and also of a fan current with a field ($x,y-jz,-2z$) \cite{pontin05a}.  Then he went on to conduct (with Klaus Galsgaard's help) fully MHD resistive simulations of non-null reconnection, which verify many of our speculations, in particular that a unique flux velocity does not exist, so that two flux velocities are needed to describe field-line motion \cite{pontin05b}. Later, we summarised the results of a series of simulations of null-point reconnection and came up with a new classification of null-point regimes, inspired by the computational results \cite{priest09a}, which includes spine-fan reconnection, torsional spine reconnection and torsional fan reconnection.  Spine-fan reconnection is the most common type and can occur during a flare above a separatrix dome \cite{pontin13a}.

Antonia started by clarifying the notion of diffusion in 1D, 2D and 3D \cite{wilmot05a}, and she went on to find fully dynamic analytical solutions to the resistive MHD equations for steady 3D reconnection in the absence of a null point \cite{wilmot06b,wilmot09c}. They consist of a sum of ideal and non-ideal solutions, with the latter exhibiting counter-rotating flows. She also gave a model for {\it flux tube disconnection}, an example of non-null reconnection.

Since then the Dundee group has been making important advances on the relaxation of braided magnetic fields \cite{wilmot09b,wilmot10a,pontin11a} and on new topological constraints during reconnection in addition to magnetic helicity conservation \cite{yeates11a,yeates13}.

\subsection{Dana Longcope -- and Coronal Heating}\label{sec7.5}

\begin{figure}
   \centering
   \includegraphics[width = 11.8cm]{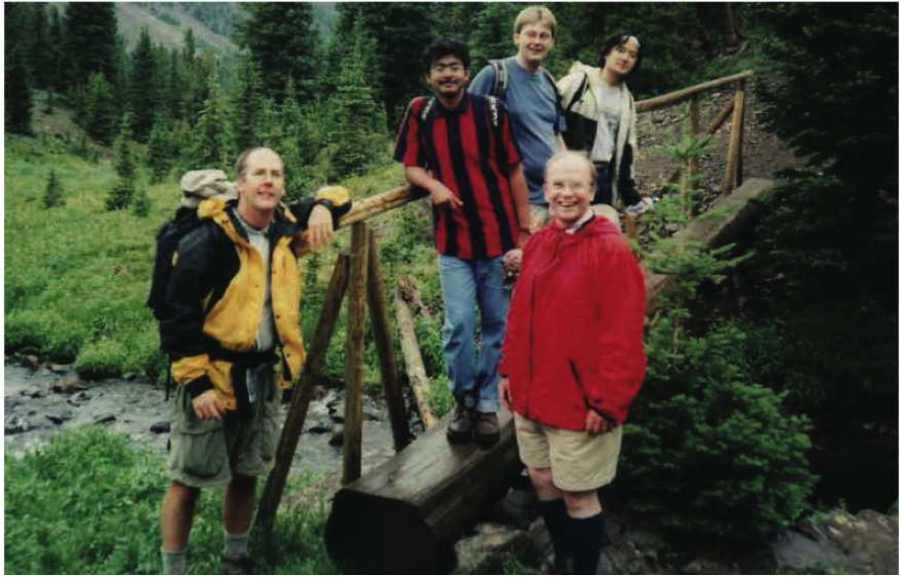}
   \caption{Walking near Bozeman in summer 2002 with Dana Longcope, Dibyendu Nandi, Stephane Regnier and Tetsuya Magara.}\label{fig21}
\end{figure}

Collaboration with Dana has been varied and great fun, starting over 10 years ago (Figure \ref{fig21}).  With Dan Brown we calculated the distribution of nulls in the solar atmosphere when the photospheric fields are random \cite{longcope03}.  The density of nulls decreases with height and depends on the difference in numbers of positive and negative fragments on the base.  We also developed with Slava Titov a concept of coronal heating by {\it binary reconnection} \cite{priest03a}, which suggests that the relative motions of myriads of magnetic fragments in the solar surface are likely to drive reconnection and therefore heating among the magnetic field lines that spread from these fragments into the solar corona. The fundamental mechanism is then one of ``binary reconnection" due to the motion of a given magnetic source relative to its nearest neighbour. The heating is due to several effects: (i) 3D reconnection of field lines that start out joining the two sources and end up joining the largest source to other more distant sources; (ii) viscous or resistive damping of the waves that are emitted by the sources as their relative orientation rotates; and (iii) relaxation of the nonlinear force-free fields that join the two sources and are built up by the relative motion of the sources.

With Jean Heyvaerts \cite{priest05a} we realised that slow photospheric motions  will in general  generate nonsingular volume currents in the overlying coronal configuration, as well as singular currents on separatrices and  separators. We derived the basic properties of current sheets and compared energy storage and heating at separators and separatrices using reduced MHD to model coronal loops that are much longer than they are wide. The result was that coronal heating is of comparable importance at separatrices and separators.

Traditionally, the waves and reconnection communities have often worked independently, but we tried to bridge the gap by calculating the fast-mode waves that are launched by sudden dissipation of a current sheet \cite{longcope07b}. The current propagates away as a sheath moving at the local Alfv\'en speed. A current density peak remains at the X-point producing a steady electric field independent of the resistivity, with reconnection driven by the external wave propagation. The majority of the magnetic energy stored by the initial current sheet is converted into kinetic energy far from the reconnection site.

\subsection{Slava Titov -- Two- and Three-Dimensional Reconnection}\label{sec7.6}

\begin{figure}
   \centering
   \includegraphics[width = 11cm]{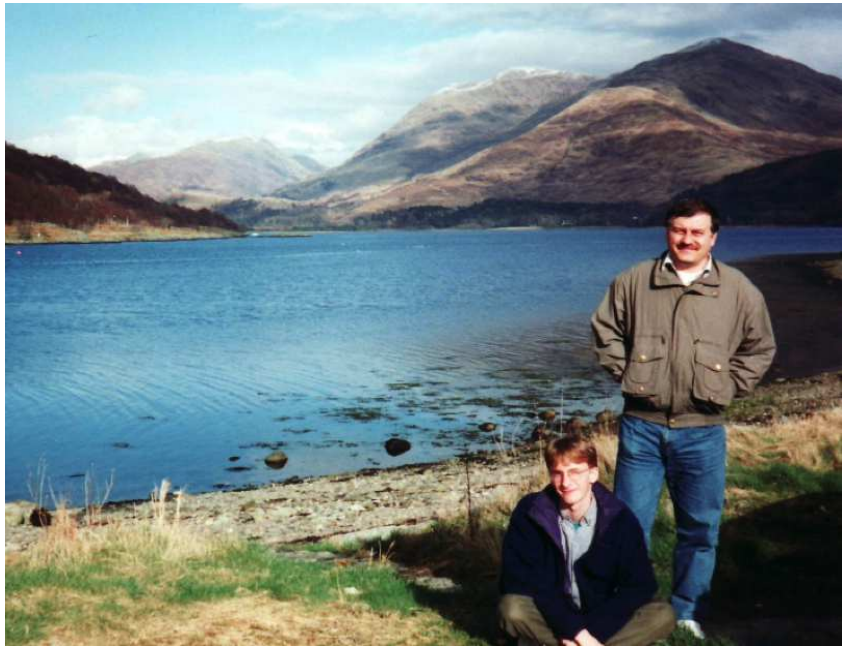}
   \caption{Slava Titov with my son Matthew visiting us in Scotland, 1992.}\label{fig22}
\end{figure}

Slava is another absolutely brilliant theorist with whom I have been privileged to interact (Figure  \ref{fig22}). After an enthusiastic letter of support from Boris Somov, I invited him to visit St Andrews in 1992, and we immediately got on really well, with our similar approach to MHD.  He has a superb classical Russian training in physics and we certainly resonate when discussing science problems.
Slava visited St Andrews on several occasions and worked in Gunnar's Topological Group in Bochum for 6 years. I was delighted when he was appointed to a long-term position with Science Applications International Corporation in San Diego, California in 2004, moving to Predictive Science Inc. in 2008.

Slava came up with an innovative and elegant new solution for the way an X-point collapses \cite{titov93b,priest95c} and for steady X-point reconnection \cite{priest94a,titov97a}. Then with Pascal's help we worked out useful conditions for the appearance of bald patches at the solar surface \cite{titov93a}. One major advance was to give the first full discussion of reconnection at 3D null points and separators \cite{priest96a} and another was to find exact solutions for reconnective annihilation \cite{priest00}.  Over these years, Slava also contributed in crucial ways to ten PhD project papers on X-ray bright points, prominences, magnetic topology and chaos.  More recently, he has produced a deep and potentially important suggestion for {\it slip-squashing factors} as a measure of 3D reconnection \cite{titov09}.

\section{Other Key Students}\label{sec8}

\begin{figure}
   \centering
   \includegraphics[width = 8cm]{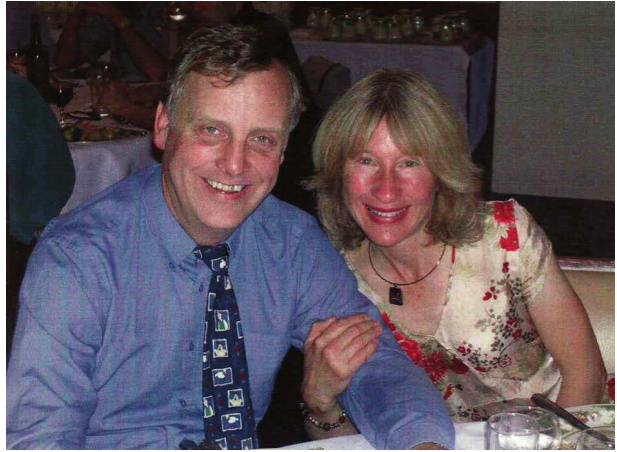}
   \caption{Alan Hood with his wife Bea at the Sept 2003 Reconnection Conference in St Andrews.}\label{fig23}
\end{figure}

Many key research students and postdocs have been mentioned already, but some whose work has not yet been described are as follows. Alan Hood has of course been a star member of the group for a long time (Figure  \ref{fig23}). He was the best student in his year as an undergraduate, so I thought ``Why don't I use this opportunity of having a really bright student to attempt a tough problem on something I  know very little about, so that together we can learn the subject?" so that is just what we did, and it worked very much as a collaboration from the start.   As a side problem, we solved the energy balance equation for a coronal loop and discovered thermal non-equilibrium when certain boundary conditions are adopted \cite{hood79a}, but the main topic was to undertake MHD stability calculations.  I had read a paper by Ulrich Anzer, suggesting that coronal flux tubes are always unstable to kinking, which I thought sounded fishy, so we set to work together going through the classical papers on MHD instability by Newcomb, Kruskal, Shafranov, Suydam and others from the laboratory plasma community.  

We realised that photospheric line-tying would be a stabilising feature and would dominate when a flux tube is weakly twisted, so we included line-tying in the analysis. The result was that for  a force-free field, the critical twist for kink instability is 3.3$\pi$ and that for magnetostatic loops it depends on the profile but typically lies between 2$\pi$ and 6$\pi$.  It is a possible cause for the eruption of a prominence in an eruptive solar flare.  Later, we refined the analysis and applied it to coronal arcades \cite{hood80} and looked into the possibility of thermal nonequilibrium for small flares \cite{hood81a}. Alan has since developed into a leading expert on instabilities, but also works on a wide range of other topics such as equilibria and emerging flux.

 Two years after Alan Hood began, another brilliant student (Peter Cargill) started his PhD with me,  and so we decided to work on quite different problems from Alan. There was at the time a lot of interest in static coronal loops (following the classic Rosner, Tucker and Vaiana paper of 1978), on which I had worked with Alan, and I had also read a paper on photospheric Evershed flow by Meyer and Schmidt (1968).  So a natural next problem was to tackle the nature of siphon flows in coronal loops, both isothermal and adiabatic, and in loops that are converging or diverging. We found a wide range of subsonic flows and flows with shocks in them \cite{cargill80a}.  

I had also read a paper by Kopp and Pneuman (1976) on the heating of flare loops, which suggested that, when a flux tube is blown open and then reconnects during a flare, it is heated by  a gas-dynamic shock propagating down the new tube from its summit. We realised that when this kinematic approach is replaced by a full dynamic approach, their shocks would be replaced by  MHD slow-mode shocks slowly propagating upwards with the reconnection point. We were able to show that such shocks can account for the observed temperatures of 5--10 MK or more and upward loop speeds of 1--10 km s$^{-1}$ \cite{cargill82a,cargill83}. 

We also suggested that the flare loops should have a cusp-like shape at their summits, and so we were delighted when this prediction was confirmed by \textit{Yohkoh} observations 10 years later \cite{tsuneta92a}. It naturally led on to numerical experiments by Terry Forbes on the same problem and to further refinements of the model in which a fast-mode shock appears below the reconnection outflow \cite{forbes83a} and the slow-mode shocks split into a conduction front and an isothermal shock \cite{forbes86b}.  In the early days, when it was thought that all the energy release occurs in the impulsive phase, these loops were called {\it post-flare loops}, but it is better to refer to them as {\it ``post"-flare loops} or, even better, just {\it flare loops}, since the energy conversion by reconnection continues throughout the main phase, often for one or even two days.  Peter then moved in 1982 to America as a postdoc in Boulder and positions at Maryland University and the Naval Research Lab. in Washington, during which he  broadened his  interests into space plasma physics and collisionless theory. It was great when he moved back to the UK to Imperial College in 1996. More recently, he retired to a wee town north of Dundee but is still very active and thankfully comes in to the group every week for a lively discussion and the group seminar.

\begin{figure}
   \centering
   \includegraphics[width = 11.8cm]{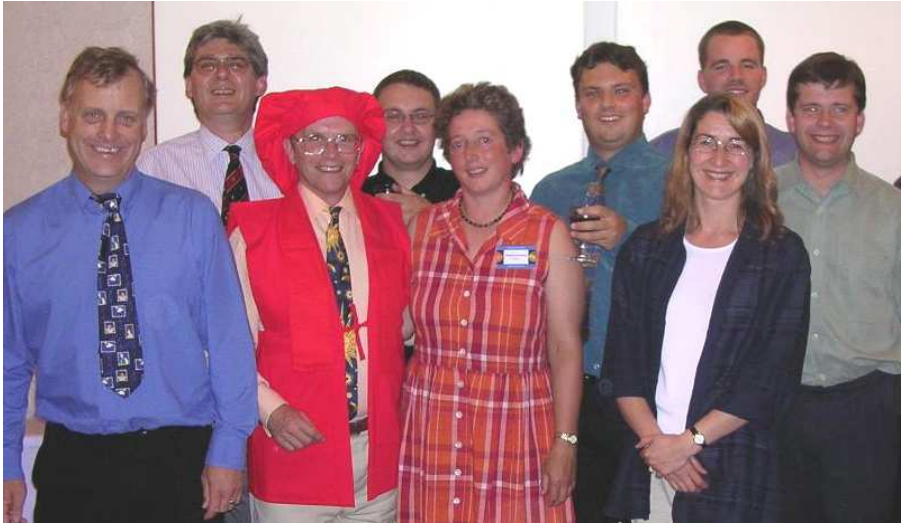}
   \caption{Some of my former research students at the September 2003 Reconnection Conference in St Andrews (my 60th birthday): Alan Hood, Peter Cargill, Colin Beveridge, Philippa Browning, Dan Brown, Moira Jardine, David Pontin and Duncan Mackay.  I am wearing one of my birthday presents (see Figure 12).}\label{fig24}
\end{figure}
Other bright students who were a joy to work with (Figure  \ref{fig24}) include Philippa Browning, who explored the basic properties of flux tubes \cite{browning82,browning83,browning84a} and went on to study coronal heating by Taylor-Heyvaerts relaxation in  coronal arcades and a series of flux-tubes \cite{browning86b,browning86c}. Later she helped apply the technique to spheromacs \cite{dixon90} and became an expert on relaxation in laboratory plasma machines.  Another such student was Moira Jardine, who did some basic research on reconnection with me \cite{jardine88a,jardine89a} and then became an authority on stellar magnetic fields, before returning to St Andrews (now a full professor in the astronomy department). A third was Dan Brown, who arrived from a master's course in nonlinear mathematics at Bath, and so I decided it would be good to tackle another tough problem about which I knew little, namely, the topology of coronal magnetic fields. To start with, Dan categorised the eight possible topologies arising from three photospheric sources, as well as the local separator, global separator and global spine bifurcations from one state to another \cite{brown99a}. We also studied the topological behaviour of separators and null points \cite{brown99b,brown01} and with Colin Beveridge the two-bipole and four-source systems  \cite{beveridge02,beveridge04}.  Dan then moved on to positions at Aberystwyth and Preston.

Two fantastic students who stayed on to take permanent positions in St Andrews are Clare Parnell and Duncan Mackay, both of whom are highly independent and original thinkers and who have their own groups of student and postdocs, so I have a ``grandfatherly" interest in them!  X-ray bright points always occur above opposite-polarity photospheric magnetic fragments, and so it was natural to suggest they are due to emerging flux. But then Karen Harvey discovered in 1984--5 that two thirds of them lie over cancelling magnetic fragments, so Clare and I decided to try and explain this.  After describing the basic features \cite{priest94b}, Clare produced a detailed model \cite{parnell94a,parnell95} and compared the potential magnetic field with the X-ray structure \cite{parnell94b}.  She also worked out the generic structure for linear 3D null points, even though it is at first sight a nine-parameter system \cite{parnell96} and then went on to discuss the way such null points tend to collapse \cite{parnell97}. Later, she developed an expertise on computational MHD and has made fundamental discoveries about the nature of the photospheric field, the magnetic carpet and the structure and behaviour of separators and 3D reconnection.

\begin{figure}
   \centering
   \includegraphics[width = 10cm]{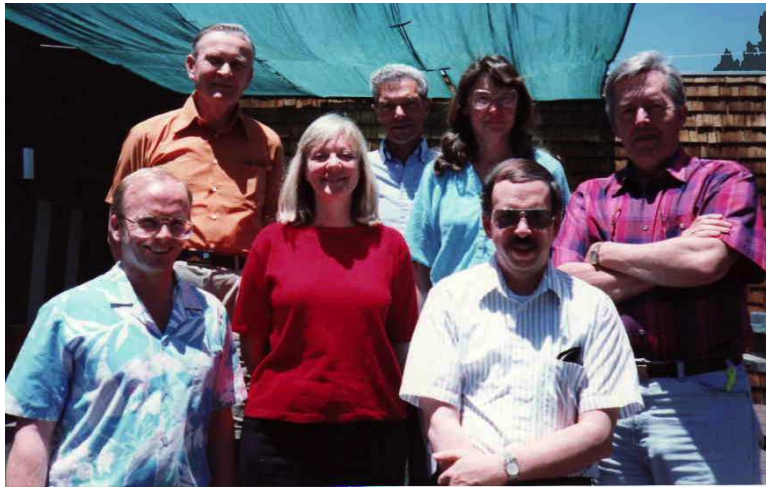}
   \caption{A meeting of the PROM group in July 1994: Vic Gaizauskas, Karen Harvey, Jack Zirker, Sara Martin, Terry Forbes and Oddbj\o rn Engvold.}\label{fig25}
\end{figure}
While Duncan Mackay was a student, we were active members of the PROM network, a small  collection of researchers led by Sara Martin who are interested in prominences (Figure  \ref{fig25}). We met annually and had highly stimulating free-flowing discussions about the latest observations and theories for prominences. This helped inform and promote a long-term interest of mine in these mysterious structures, and Duncan decided with my help to develop a series of models for filament channels and prominence magnetic fields \cite{mackay96,mackay97a,mackay97b,mackay98,mackay99}. After that he set up models for global magnetic flux transport over the surface of the Sun \cite{mackay02}. Since then he has linked up with Aad van Ballegooijen (one of the most brilliant minds in solar physics in my opinion) to model with Anthony Yeates global coronal nonlinear force-free fields that are driven by footpoint evolution. Comparison with observations was stunning -- they managed to predict most of the locations for prominence formation (large-scale twisted flux tubes) and for prominence eruption.

\section{Other Aspects of My Life}\label{sec9}

\begin{figure}
   \centering
   \includegraphics[width = 11cm]{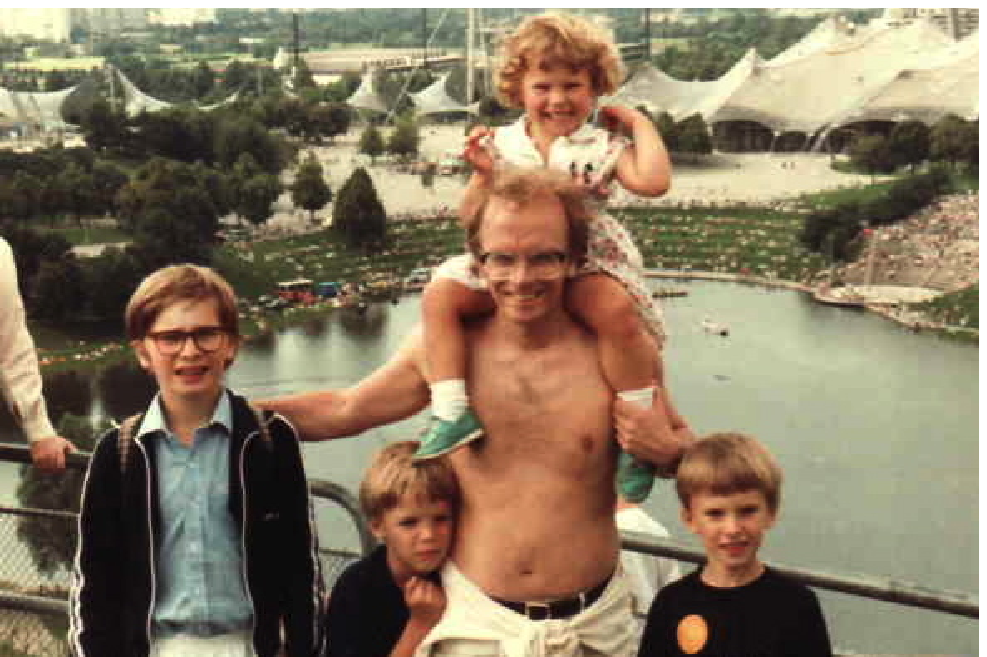}
    \includegraphics[width = 11cm]{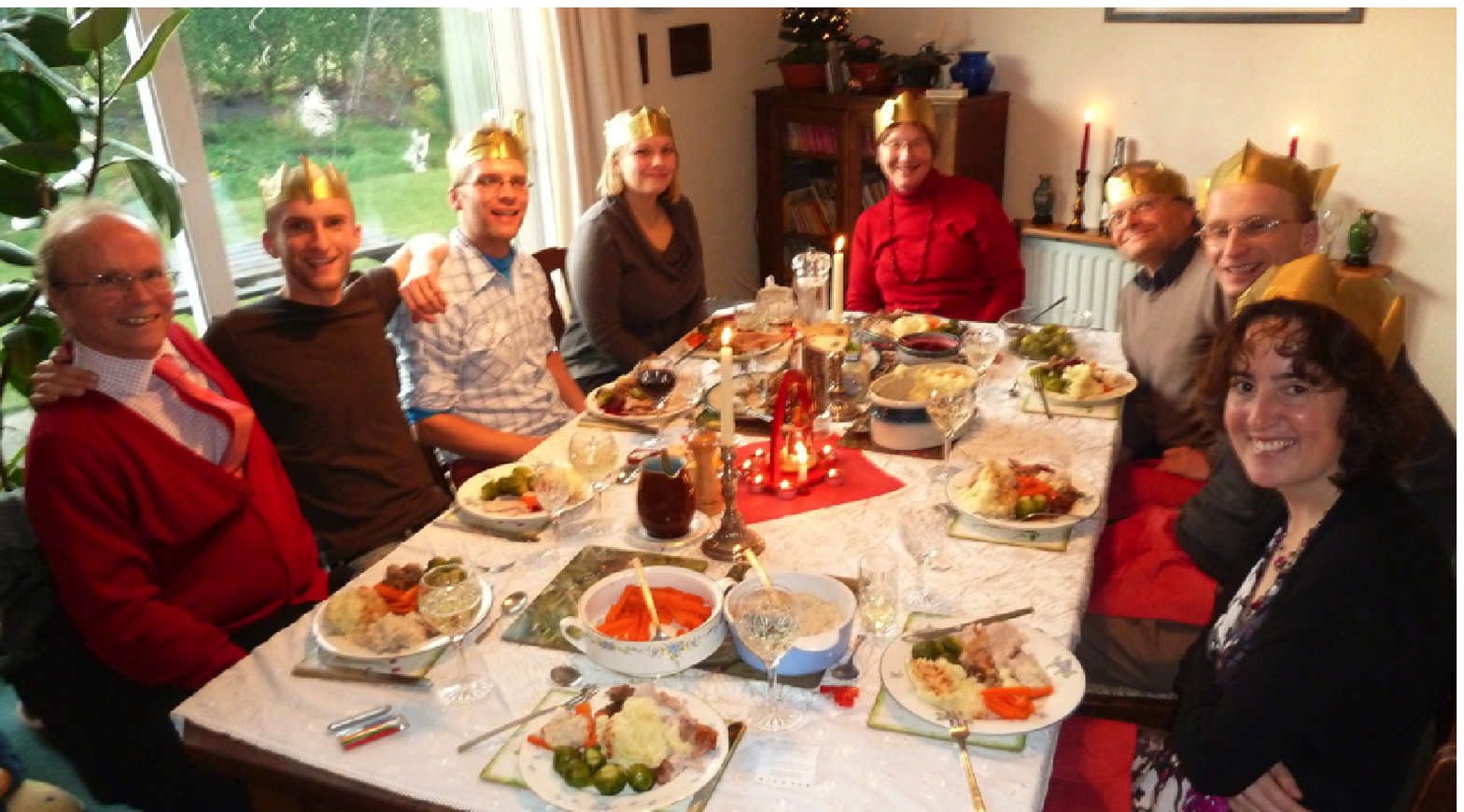}
   \caption{The family (a) in Munich in 1982 and (b) in St Andrews at Christmas 2011 (Matthew, David, Naomi, Clare, Gerry, Andrew and his wife Cathy.}\label{fig26}
\end{figure}

Many other aspects of my life are important to me in addition to research and teaching.  Foremost of these is my family (Figure  \ref{fig26}), who have been highly supportive and tolerant -- a standard reply to ``Oh is daddy working again?" is ``No, I've just been playing!"  The greatest joy of all is to have shared my life with Clare, who has always been a calm, kind and loving influence.  Our four children, a son (Andrew), twin-sons (David and Matthew) and then a wee daughter (Naomi), have given us a huge pleasure over the years. Time is the most important thing that a father can give a child, and I spent a lot of it with them each when they were young. We always cherish spending Christmas and summer holidays all together.

Another aspect is Christianity. My basic approach is that we know a miniscule amount about the universe and a miniscule amount about God and so have no right to be arrogant about either.  Therefore, valuing other people's insights on science and faith is important.  Just as I cannot prove a theory is correct, but can only say whether or not it is consistent with the observations, so there is no way I can prove that God exists, but can only say whether his existence is consistent with my experience. For me personally, it is, and so I am living my life on that assumption until I find otherwise. 

To me there are many parallels between life as a scientist and as a Christian, and so I have always been interested in the whole area of science and religion.  Being a scientist means being open to creativity and new ideas, having leaps of faith, expecting your ideas will change, and being part of a community.  It often fills you with a sense of  beauty,   wonder and humility.  My experience is that being a Christian has just the same elements.

Other activities I enjoy have been mentioned in Section \ref{sec2}, namely, music and exercise, both of which are mentally as well as physically relaxing. I also enjoy attending theatre or an art gallery when there is time off while visiting a big city such as London.

I hold the James Gregory chair in St Andrews, named after a great scientist, who was the first Regius Professor of Mathematics in St Andrews, appointed by King Charles II in 1668 at the age of 30. He moved to Edinburgh after 6 years but died a year later at the age of 36.  In his short life he made major contributions. He laid down one of the world's first {\it astronomical meridian lines} in St Andrews, invented the {\it gregorian telescope}, and discovered the {\it diffraction grating} after shining light through a seagull's feather that he had perhaps picked up after a walk along the beach.  He was also one of the three inventors of {\it calculus} (with Newton and Leibniz) and taught calculus in St Andrews 100 years before Cambridge. He discovered Taylor series (40 years before Taylor) and wrote the first text book on calculus, having proved many of the fundamental results about integration as well as the fundamental theorem of calculus.  So you will not be surprised to hear that I have been spending some time trying to commemorate him in various ways around the town so that his achievements become better known.

\section{Conclusion}\label{sec10}

My own contribution to solar MHD is a series of small additions to understanding, in which I have been greatly privileged to share ideas with a series of wonderful people mentioned here and many others not mentioned.  This has been within a lively and vibrant international environment, with ideas being freely tossed around and shared. There have been healthy close links between observation, theory and computational experiment, and so I hope they continue, since such links are crucial for the future.   In future, I expect many models to become more realistic by going beyond single-fluid resistive MHD.

At the moment, I am working on a series of interesting topics: with Clare Parnell and her students on separator reconnection and coronal topology; with David Pontin on 3D reconnection; with Miho Janvier on magnetic helicity conservation in coronal eruptions; with Fernando Moreno Insertis on prominence tornadoes; with Hubert Baty and Terry Forbes on fast reconnection; and with Vasilis Archontis and  Duncan Mackay on magnetohydrostatic extrapolation, so there are plenty of new ideas to keep me occupied.
 
\begin{figure}
   \centering
   \includegraphics[width = 9cm]{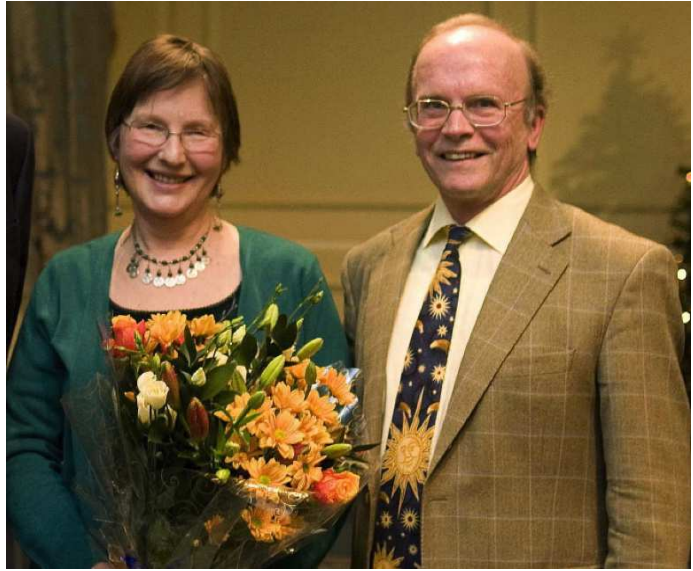}
   \caption{With Clare at my ``retirement" from the University in November, 2009.}\label{fig28}
\end{figure}
For me, it is important to help foster a sense of community, in which all are valued and all play a part, and in which we all support each other.  
It is also essential to allow time to ponder, without rush or stress, so that creativity and intuition may be encouraged. We need to counter the current climate of too much stress, expectation, administration and reporting and let the pendulum swing back to a calmer and more creative state.  In all this, quality and personal relations are more important than quantity and lack of appreciation.

I have been a highly fortunate person to have found a sense of fun in my work and to have been able to learn from so many talented colleagues across the world.  The group in St Andrews have represented a wonderfully supportive, lively and creative environment for many years, especially Bernie, Alan, Thomas, Clare, Duncan and Ineke. However, without the continued love and care of my wife, Clare, very little would have been achieved (Figure  \ref{fig28}).

\begin{acknowledgements} 
I am delighted to thank all my friends and colleagues across the world for their friendship and help, as well  PPARC, STFC and the Leverhulme Trust for financial support.
\end{acknowledgements}

\end{document}